\documentclass[useAMS,usenatbib]{mn2e}
\usepackage{amssymb}
\usepackage{natbib}
\usepackage{times}
\usepackage{graphics,epsfig}
\usepackage{graphicx}
\usepackage{amsmath}
\usepackage{amssymb}
\usepackage{comment}
\usepackage{supertabular}

\usepackage{graphicx}
\usepackage{amsmath}
\usepackage{color}

\newcommand{\kms}{km~s$^{-1}$}
\newcommand{\arcs}{$^{\prime\prime}$}

\newcommand{\hh}{\ensuremath{\rm H_{2}}}

\newcommand{\mspcc}{\ensuremath{\rm M_{\odot} pc^{-3} }}
\newcommand{\HI}{H{\sc i }}

\title {Molecular scale height in spiral galaxies}
\author [N. N. Patra]{	Narendra Nath Patra$^{1,2}$ \thanks {E-mail: narendra@rri.res.in} \\
	$^{1}$ Raman Research Institute, C. V. Raman Avenue, Sadashivanagar, Bengaluru 560080, India\\
	$^{2}$ National Centre for Radio Astrophysics, Tata Institute of Fundamental Research, Pune University campus, Pune 411 007, India\\
}
\date {}
\begin{document}
\maketitle

\begin{abstract}

Having to have low thermal energy, the molecular gas in galaxies is expected to settle in a thin disc near the midplane. However, contradicting this understanding, recent studies have revealed considerably thick molecular discs in nearby spiral galaxies. To understand this apparent discrepancy, we theoretically model the molecular discs in a sample of eight nearby spiral galaxies and estimate their molecular scale heights (Half Width at Half Maxima (HWHM)). We assume that the baryonic discs are in vertical hydrostatic equilibrium under their mutual gravity in the external force field of the dark matter halo. We set up the joint Poisson's-Boltzman equation of hydrostatic equilibrium and numerically solve it to obtain the three-dimensional molecular gas distribution and determine the scale heights in our sample galaxies. We find that the scale heights follow a universal exponential law with a scale length of $0.46 \pm 0.01 \ r_{25}$. The molecular scale heights in our sample galaxies are found to vary between 50-200 pc depending on the galaxy and radius. Using the density solutions, we build dynamical models of the molecular discs and produce molecular column density maps. These model maps found to match to the observed ones reasonably well. We further incline the dynamical models to an inclination of 90$^o$ to estimate the expected observed thickness of the molecular discs. Interestingly it is found that at edge-on orientation, our sample galaxies under hydrostatic assumption can easily produce a few kpc thick observable molecular disc.

\end{abstract}


\section{Introduction}

The molecular gas in galaxies plays a crucial role in transforming the gas into stars. As the molecular gas is the immediate precursor to the star formation, this component of the Interstellar medium (ISM) has a potentially significant influence on galaxy formation and evolution \citep{myers86,shu87,scoville89}. Not only that, but dynamically also this component is critical. For example, in the inner part of spiral galaxies, the molecular gas dominates the ISM and influence the dynamics at the central region \citep{leroy09a,richards18}. However, in spite of its enormous importance, accurate three-dimensional distributions of molecular gas in galaxies are not well understood \citep{cohen86,hunter97,dame01,sawada04}. The lack of spatial resolution and sensitivity limits the detection of molecular clouds even in the nearby galaxies \citep[see, e.g.,][and references therein]{schruba17}. Whereas, with a larger beam, the line-of-sight integration effects severely restrict one to extract the vertical distribution of the molecular gas even for an edge-on galaxy.  

Traditionally, it is thought that the molecular component is a dynamically cold component which settles in a thin disc near the midplane with small vertical thickness \citep[see, e.g.,][]{nakanishi06}. Consequently, the vertical velocity dispersion of the molecular gas is also expected to be small ($\sim$ a few \kms). However, many recent observational studies indicate the existence of a diffuse low-density molecular gas in galaxies which can extend to a much larger height from the midplane \citep{garcia-burillo92,pety13,calduprimo13,mogotsi16}. This component is thought to be well mixed with the atomic gas and might be a part of the same dynamical component. However, currently, we lack a theoretical understanding of the origin and sustenance of this component, which makes it interesting to investigate the distribution of the molecular gas in galaxies, especially in the vertical direction.

Observing out of the plane molecular gas distribution is challenging in external galaxies due to line-of-sight integration effects. A detailed study of the three-dimensional distribution of the molecular gas is only possible observationally for the MilkyWay \citep{grabelsky87,grabelsky88,bronfman88,rosolowsky06,nakanishi06,rice16}. For example, using the observations made by the Columbia University 1.2 m telescope at Cerro Tololo, \citet{grabelsky87} reported for the first time out of the plane distribution of the molecular gas outside the solar circle. \citet{wouterloot90} used IRAS point source catalogue and $^{12}$CO observations to estimate the distribution of the molecular gas in the outer parts of the Galaxy. Later, using the data from the Columbia $^{12}$CO ($ J = 1 \rightarrow 0$) survey \citep{dame01} \citet{nakanishi06} produced a three-dimensional map of the molecular gas for the entire MilkyWay. They found that the molecular scale height in the Galaxy varies between $\sim 50-200$ pc as one moves from the centre to a radius of $\sim$ 10 kpc. However, many of these studies are severely hindered by distance ambiguities and high opacity of the molecular clouds \citep{grabelsky87,sanders84,scoville87,bronfman88}. 

To acquire a broader picture of the molecular gas distribution in galaxies, one needs to extend the sample beyond the MilkyWay. However, for external galaxies, it is extremely hard to resolve out individual clouds due to the limited spatial resolutions of the present day telescopes. It is only very recently; high-resolution studies could be possible for a handful of very nearby galaxies with the advanced interferometers like the Atacama Large Millimetre Array (ALMA) \citep{schruba17}. Hence, observationally, it is difficult to estimate the three-dimensional distribution of the molecular gas directly and consecutively measure the thickness of the molecular discs. Nonetheless, observations of the molecular discs in edge-on galaxies found a much higher thickness of the molecular discs than what is presumed before. Not only that, recent observations of the molecular gas velocity dispersions (which is an indirect measure of the thickness of a molecular disc) in nearby spiral galaxies reconfirm the existence of a thick molecular disc in external galaxies \citep{garcia-burillo92,pety13,mogotsi16}.

To theoretically understand the thickness of the molecular discs and compute the three-dimensional distribution of the molecular gas, we assume the baryonic discs in a galaxy to be in vertical hydrostatic equilibrium under their mutual gravity; in the external force field of the dark matter halo. With this assumption, we set up and numerically solve the joint Poisson's-Boltzman equation to theoretically estimate the distribution of the molecular gas in galaxies. We used this approach in our earlier paper \citep{patra18a} to solve the hydrostatic equilibrium equation and estimate the molecular gas distribution in the galaxy NGC 7331. We found that in NGC 7331, our modelled molecular disc very well matches with the observation. This motivates us to use this theoretical approach to solve the hydrostatic equilibrium equation to estimate the distribution of molecular gas in a larger sample of external galaxies. As mentioned earlier, the determination of the molecular gas distribution observationally or theoretically is attempted only for a handful of galaxies, in this paper, we try to estimate the distribution of molecular gas and hence the thickness of the molecular discs in a sample of eight nearby spiral galaxies.

\section{Sample}

We select our sample galaxies from the HERA CO-Line Extragalactic Survey (HERACLE) \citep{leroy09a} in which 18 nearby galaxies were observed in CO using the 30-m IRAM telescope. To solve the hydrostatic equilibrium equation, we are required with the deprojected surface densities of different baryonic components (e.g., stars, \HI~and \hh), observed rotation curves and the mass-models of the galaxies. The molecular surface density profiles are obtained from the HERACLE survey whereas The \HI~Nearby Galaxy Survey (THINGS) \citep{walter08} data provide with the atomic gas surface densities, rotation curves and the mass models. 18 galaxies were observed as part of the HERACLE survey, out of which four galaxies were not detected in CO (Holmberg-I, Holmberg-II, IC 2574 and DDO 154). We considered the rest 14 galaxies as potential sample galaxies for our analysis. However, out of these 14 galaxies, rotation curves could not be derived for four galaxies (NGC 628, NGC 3148, NGC 3351 and NGC 4214) due to their low inclinations, $i \lesssim 40^o$. Hence, we exclude these four galaxies for further consideration. However, for one galaxy in the parent sample, NGC 6946 a reliable rotation curve could be derived in spite of it having an inclination, $i \sim 33^o$ as it had a significant number of resolution elements across its major axis \citep[see][for more details]{deblok08}. Hence, we include NGC 6946 in our sample. This leaves us with a sample of 10 potential galaxies having all the surface densities and rotation curves. Further, for two galaxies (NGC 2903 and NGC 4736) mass modelling could not be done due to complex kinematic signatures in their \HI~data. For example, in NGC 2903, due to the presence of a bar, a significant non-linear motion is induced leading to a non-reliable mass modelling of the galaxy \citep[see, e.g., ][]{deblok08}. The rotation curve of NGC 4736 on the other hand, found to be very complex with the presence of strong non-linear motions \citep{trachternach08}. This, in turn, makes it nearly impossible to determine the dark matter distribution in this galaxy using just the observed rotation curve alone \citep{deblok08}. We exclude these two galaxies further from our potential sample galaxies. These considerations finally leave us with eight nearby spiral galaxies for which we can solve the hydrostatic equilibrium equation. In Tab.~\ref{tab:samp} we list the general properties of our sample galaxies. In column (1) we list the names of the galaxies whereas column (2) and (3)  present the distance to the galaxies and the inclination respectively. In column (4) and (5) we list the position angle of the optical disc and the $r_{25}$ radius respectively. The data presented in Tab.~\ref{tab:samp} are taken from \citet{walter08}.

\begin{table}
\caption{Sample galaxies}
\begin{center}
\begin{tabular}{lcccc}

\hline
Name & Dist & Incl & PA & $r_{25}$\\
     &(Mpc) & ($^o$) & ($^o$) & ($^\prime$)\\
\hline
NGC 925   &  9.2   &  66  &  287  &  5.3\\
NGC 2841  &  14.1  &  74  &  154  &  5.3\\
NGC 2976  &  3.6   &  65  &  335  &  3.6\\
NGC 3198  &  13.8  &  72  &  215  &  3.2\\
NGC 3521  &  10.7  &  73  &  340  &  4.2\\
NGC 5055  &  10.1  &  59  &  102  &  5.9\\
NGC 6946  &  5.9   &  33  &  243  &  5.7\\
NGC 7331  &  14.7  &  76  &  168  &  4.6\\
\hline

\end{tabular}
\end{center}
\label{tab:samp}
\end{table}

\begin{figure*}
\begin{center}
\begin{tabular}{c}
\resizebox{1.\textwidth}{!}{\includegraphics{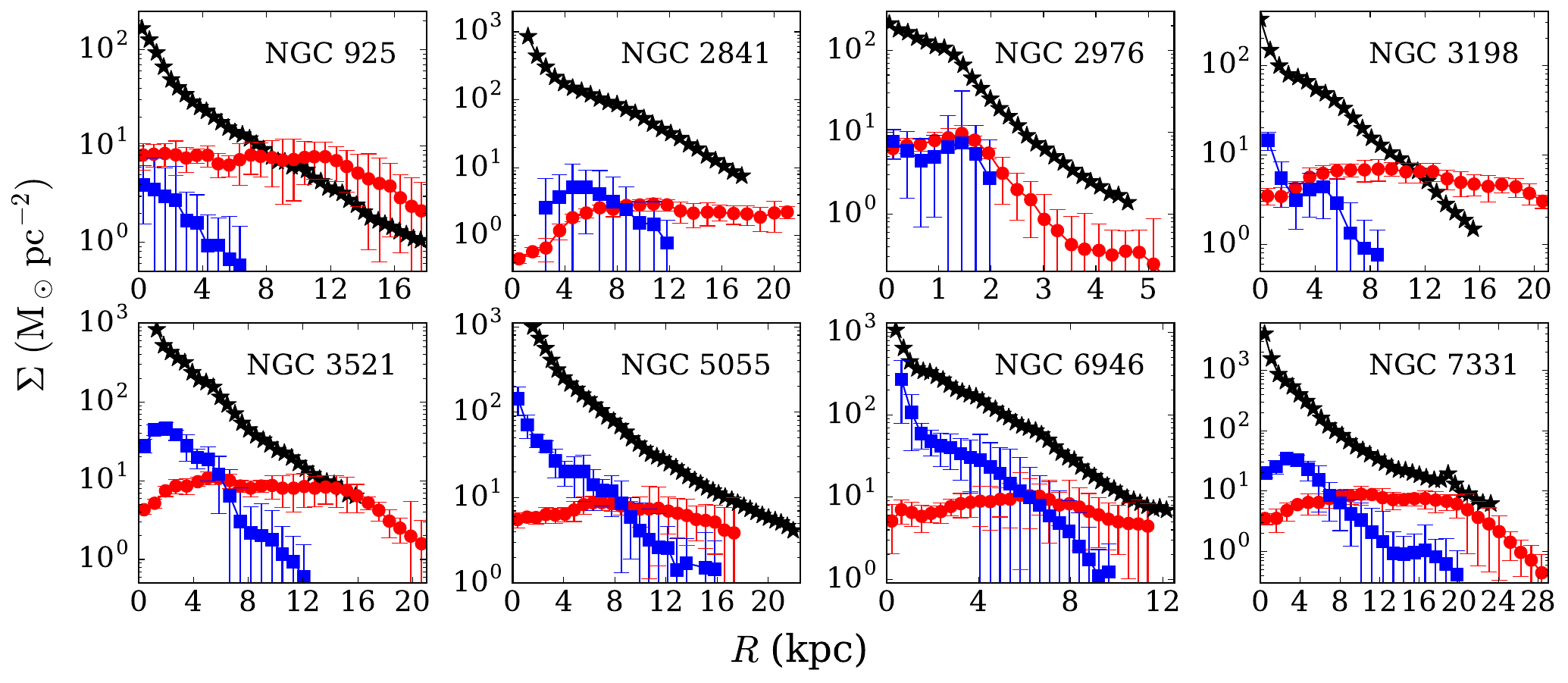}}
\end{tabular}
\end{center}
\caption{The surface density profiles of our sample galaxies. Different panels plot the surface density profiles for different galaxies as mentioned in the top right corner of the respective panels. In each plot, the black asterisks represent the stellar surface density profiles whereas the red filled circles with error bars show the \HI~surface density profiles whereas the blue squares represent the \hh~surface density profiles. The \hh~surface density profiles are taken from the HERACLE survey \citep{leroy09a} whereas the stellar and the \HI~surface density profiles were taken from \citet{leroy08}.}
\label{sden}
\end{figure*}

\section{Modelling the galactic disc}

\subsection{The equation of hydrostatic equilibrium}

To estimate the molecular distribution in our sample galaxies, we assume a galaxy to be a three-component system consists of a stellar disc, an atomic disc and a molecular disc settled under mutual gravity under the gravitational influence of the dark matter halo. Each of these discs is in vertical hydrostatic equilibrium under the balance between the gravity and the pressure. The gravity is the total gravity produced by all the components whereas the balancing pressure would be of individual components decided by their velocity dispersion. To simplify our model for mathematical convenience, we assume that all the discs are coplanar, coaxial and concentric. Not only that, but the dark matter halo would also be symmetric, and the centres of the dark matter halo and the discs coincide with each other. Under these assumptions, for an elemental volume inside the galactic disc, the Poisson's equation of hydrostatic equilibrium in cylindrical coordinate can be written as  

\begin{equation}
\label{eq1}
\frac{1}{R} \frac{\partial }{\partial R} \left( R \frac{\partial \Phi_{total}}{\partial R} \right) + \frac{\partial^2 \Phi_{total}}{\partial z^2} = 4 \pi G \left( \sum_{i=1}^{3} \rho_{i} + \rho_{h} \right)
\end{equation}

\noindent where $\Phi_{total}$ is the total gravitational potential due to all the baryonic discs and the dark matter halo. $\rho_i$ runs for the stellar, atomic and the molecular mass densities whereas $\rho_h$ denotes the mass density of the dark matter halo. Here we consider the dark matter halo to be of fixed structure (not live) determined by the observed rotation curve. For our sample galaxies, the dark matter halo parameters are extracted using THINGS survey data by \citet{deblok08}. \citet{deblok08} used both the NFW profile and the isothermal profile to describe the dark matter halos. An NFW profile \citep{navarrofrenkwhite97} can be given as 

\begin{equation}
\label{nfw}
\rho_h(R) = \frac {\rho_0}{\frac{R}{R_s} \left( 1 + \frac{R}{R_s}\right)^2}
\end{equation}

\noindent whereas an isothermal profile can be given as 

\begin{equation}
\label{eq_iso}
\rho_h(R) = \frac {\rho_0}{1 + \left(\frac{R}{R_s}\right)^2}
\end{equation}

\noindent where $\rho_0$ is the characteristic density and $R_s$ is the scale radius. These two parameters ($\rho_0$ and $R_s$) completely describe a spherically symmetric dark matter halo.

The baryons in the galactic discs can be assumed to be an ideal gas for which the pressure gradient would be balanced by the gradient in the gravitational potential under hydrostatic equilibrium. 

\begin{equation}
\label{eq4}
\frac{\partial }{\partial z} \left(\rho_i {\langle {\sigma}_z^2 \rangle}_i \right) + \rho_i \frac{\partial \Phi_{total}}{\partial z} = 0
\end{equation}

\noindent where ${\langle {\sigma}_z \rangle}_i$ is the vertical velocity dispersion of the $i^{th}$ component. 

\noindent Using Eq. (1) and (4) we get,

\begin{equation}
\label{eq5}
\begin{split}
{\langle {\sigma}_z^2 \rangle}_i \frac{\partial}{\partial z} \left( \frac{1}{\rho_i} \frac{\partial \rho_i}{\partial z} \right) &= \\ 
&-4 \pi G \left( \rho_s + \rho_{HI} + \rho_{H_2} + \rho_h \right)\\
&+ \frac{1}{R} \frac{\partial}{\partial R} \left( R \frac{\partial \Phi_{total}}{\partial R} \right)
\end{split}
\end{equation}

\noindent where $\rho_s$, $\rho_{HI}$ and $\rho_{H_2}$ are the mass densities of stars, atomic gasses and the molecular gasses respectively.

\noindent The above equation can be further simplified by using the fact \citep[see][for more details]{banerjee10},

\begin{equation}
{\left( R \frac{\partial \Phi_{total}}{\partial R} \right)}_{R,z} = {(v_{rot}^2)}_{R,z}
\end{equation}

\noindent where ${(v_{rot})}_{R,z}$ is the rotation velocity. If one neglect the variation of the rotation velocity in the $z$ direction then the ${(v_{rot})}_{R,z}$ can be replaced by the observed rotation curve, $v_{rot}$.
 
\noindent Thus Eq.~\ref{eq5} reduces to

\begin{equation}
\begin{split}
{\langle {\sigma}_z^2 \rangle}_i \frac{\partial}{\partial z} \left( \frac{1}{\rho_i} \frac{\partial \rho_i}{\partial z} \right) &= \\
&-4 \pi G \left( \rho_s + \rho_{HI} + \rho_{H_2} + \rho_h \right)\\ 
&+ \frac{1}{R} \frac{\partial}{\partial R} \left( v_{rot}^2 \right)
\end{split}
\label{eq_hydro}
\end{equation}

\noindent Eq.~\ref{eq_hydro} is the final hydrostatic equilibrium equation which represents three second-order ordinary partial differential equation in the variables $\rho_s$, $\rho_{HI}$ and $\rho_{H_2}$ which are coupled to each other through the first term on the RHS. Solving this equation would provide a detailed three-dimensional density distribution of the baryonic discs.

\begin{figure*}
\begin{center}
\begin{tabular}{c}
\resizebox{1.\textwidth}{!}{\includegraphics{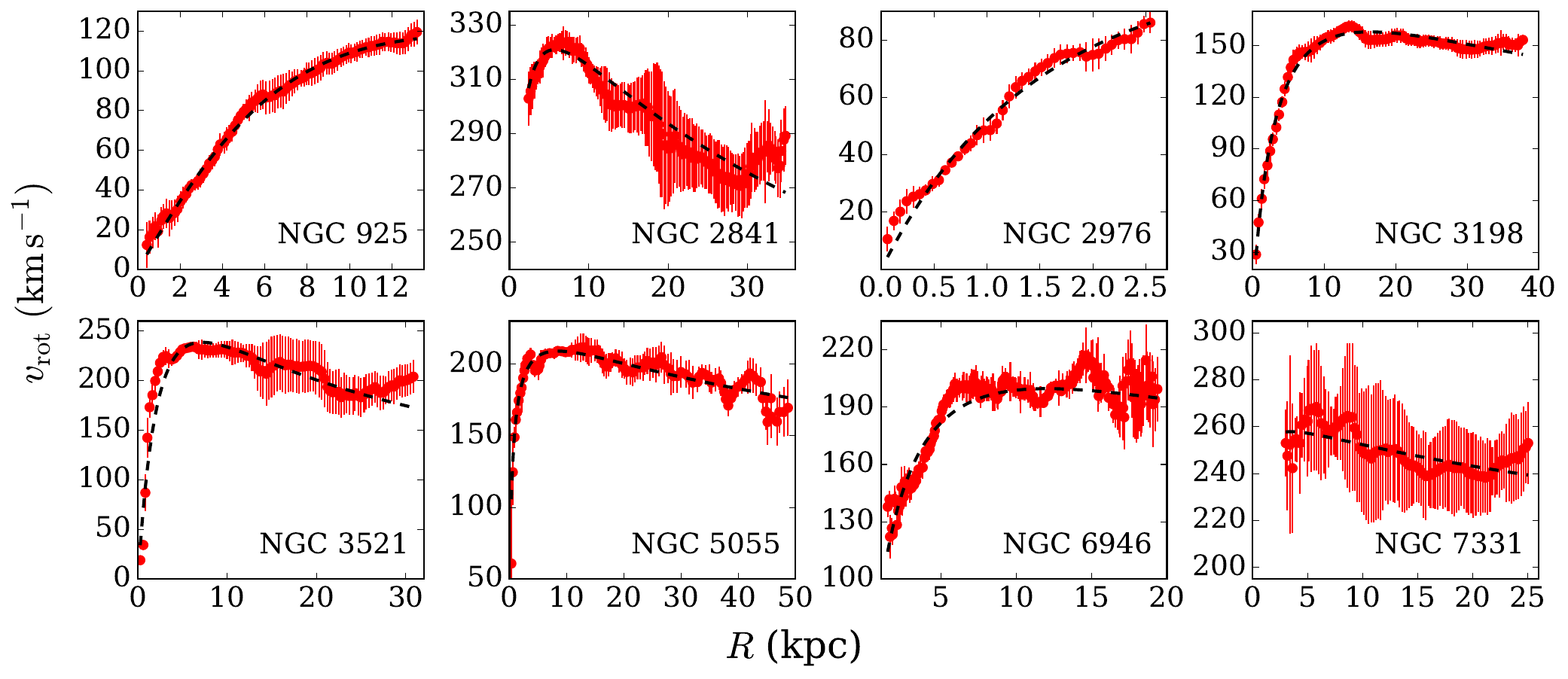}}
\end{tabular}
\end{center}
\caption{The rotation curves of our sample galaxies as extracted from the THINGS data \citep{deblok08}. For parametrisation, we fit the rotation curves with a Brandt profile. The red circles with error bars represent the rotation curves whereas the black dashed lines represent the fit to the data. The best-fit parameters are given in Tab.~\ref{tab:rotcur}.}
\label{rotcur}
\end{figure*}

\subsection{Inputs Parameters}
\label{input_params}

Solving Eq.~\ref{eq_hydro} analytically is not possible, and hence, we solve it using numerical techniques. There are a number of essential inputs which are necessary to solve the equation. Below we discuss these inputs in details.

\subsubsection{Surface densities}

In hydrostatic equilibrium, gravity balances the pressure in the vertical direction. The primary source of the gravity is the observed baryons in the galactic discs apart from the dark matter halo. Hence, the surface densities of the baryonic components are one of the primary inputs to Eq.~\ref{eq_hydro}. For our sample galaxies, the surface densities of the molecular gas are obtained using the HERACLE survey data. We adopt the molecular surface density profiles of our sample galaxies from \citet{schruba11}. \citet{schruba11} averaged the CO data in tilted rings of width 15\arcs~(taken from the THINGS survey \citep[see][for more details]{deblok08}) by shifting the spectra to a common velocity and stack them. The surface density profiles of the molecular gas are then calculated by fitting these stacked spectra. As the stacked spectra have much higher SNR, the surface density profiles obtained by this method are much more sensitive than what would be obtained using a moment zero map \citep[for example what is presented in][]{leroy09a}. It can be noted that the molecular surface density profiles thus generated extend to a larger radius than what would be seen in a regular moment map. The surface density profiles of the atomic gas are obtained by averaging the total \HI~intensity distribution in the same tilted rings \citep{schruba11}. A correction factor of 1.4 is applied to the surface density profiles of the atomic gas to account for the presence of the cosmological Helium. We adopt the stellar surface density profiles for our sample galaxies as calculated by \citet{leroy08} using 3.6 $\mu m$ data from the SINGS survey \citep{kennicutt03}. In Fig.~\ref{sden} we plot the different surface density profiles of our sample galaxies. It can be seen from the figure that the stellar and the atomic discs extend to a much larger radius than the molecular discs. Also, the stellar disc dominates the surface density in the inner parts of the galaxies, whereas, the atomic gas becomes comparable or dominant at the outer parts. These surface density profiles are used as an input to solve the hydrostatic equilibrium equation.

\subsubsection{Vertical velocity dispersion}

The next important input required to solve Eq.~\ref{eq_hydro} is the vertical velocity dispersions of individual disc components. It is a key input as it alone provides the necessary pressure to balance the gravity and can significantly influence the vertical scale heights of the baryonic discs.

We calculate the stellar velocity dispersion analytically by assuming the stellar disc to be an isothermal system in hydrostatic equilibrium \citep[see Appendix B.3 of][for more details]{leroy08}. However, recent studies of stellar velocity dispersion using IFU data suggest that the velocity dispersion obtained assuming an isothermal stellar disc is always an overestimate of the true velocity dispersion within the disc scale-length, and an underestimate outside the disc scale-length \citep[see][for more details]{mogotsi18}. Nonetheless, \citet{banerjee11} showed that the vertical velocity dispersion of the stellar component does not significantly influence the scale heights of the gaseous discs which are of our prime interest. Hence, we use the analytical formula as given by \citet{leroy08} to calculate the vertical velocity dispersions of the stellar discs in our sample galaxies.

Unlike the stellar velocity dispersion which is very hard to measure observationally, the gas velocity dispersions are relatively easy to realise through spectroscopic observations. The velocity dispersion of the atomic gas in galaxies can be studied extensively using $\rm HI-21cm$ spectroscopic observations. Earlier low-resolution \HI~studies revealed a velocity dispersion of 6-13 \kms~in the atomic discs. However, consecutive interferometric resolved studies measured the gas velocity dispersion with more precision. For example, \citet{tamburro09} used high-resolution \HI~data of THINGS galaxies to find a mean velocity dispersion of $\sim$ 10 \kms~in their atomic discs. However, these studies are based on individual spectrum which very often lacks sensitivity. On a global scale, however, spectral stacking produce \HI~spectra with much higher sensitivity and are able to detect faint broad wings which are not detectable in the high-resolution individual spectra \citep[see e.g.,][] {stilp13}. These stacked spectra resulted in somewhat higher velocity dispersions of \HI~gas than what was found earlier. \citet{calduprimo13} stacked the CO and the \HI~spectra in a sample of spiral galaxies (using the same parent sample what we have used here) to find a $\sigma_{HI}/\sigma_{co} = 1.0 \pm 0.2$ with a median $\sigma_{HI}$ of $11.9 \pm 3.9$ \kms. Later, \citet{mogotsi16} used the same set of galaxies to study the individual bright spectra and found a $\sigma_{HI}/\sigma_{co} = 1.4 \pm 0.2$ with a median $\sigma_{HI} = 11.7 \pm 2.3$ \kms~and $\sigma_{co}=7.3 \pm 1.7$ \kms. Given these results, it is reasonable to assume a velocity dispersion of $\sim$ 12 \kms~for the atomic gas. However, in the molecular disc, the dominant component has a velocity dispersion of $\sim$ 7 \kms~which is observed in both the individual spectrum and stacking analysis. The other diffuse component which is only seen in the stacked spectra has a much higher velocity dispersion of $\sim$ 12 \kms. For our analysis here, we consider the molecular disc to be a single component system with a velocity dispersion of 7 \kms~and investigate if this simple molecular disc can explain the observed thickness of the molecular discs in external galaxies.

\subsubsection{Rotation curve}

The rotation curve is another important input to the hydrostatic equation. The last term on the RHS of Eq.~\ref{eq_hydro} requires the observed rotation curve to be known. This term provides a centripetal force which effectively works like a pressure against gravity. For our sample galaxies, we use rotation curves as derived from the high-resolution \HI~data from the THINGS survey \citep{deblok08}. We plot the rotation curves of our sample galaxies in Fig.~\ref{rotcur}. As a second-order derivative is what is used in Eq.~\ref{eq_hydro}, we parametrise the rotation curves using a commonly used Brandt profile \citep{brandt60} to avoid any possible divergence. The profile can be given as 

\begin{equation}
v_{rot} (R) = \frac{V_{max}\left(R/R_{max} \right)}{\left(1/3 + 2/3 \left(\frac{R}{R_{max}}\right)^n\right)^{3/2n}}
\end{equation}

\noindent where the $V_{max}$ is the maximum rotational velocity, $R_{max}$ is the radius at which this maximum velocity occurs. $n$ is the power-law index which signifies how fast the rotation curve increases as a function of radius.

\begin{table}
\caption{Best fit parameters of the rotation curves}
\begin{center}
\begin{tabular}{lccc}

\hline
Name & $V_{max}$ & $R_{max}$ & $n$ \\
     &(\kms) & (kpc) & \\
\hline
NGC 925   & 118.0$\pm$1.4 & 16.4$\pm$0.9 & 1.80$\pm$0.10\\
NGC 2841  & 320.9$\pm$0.5 & 5.7$\pm$0.2  & 0.36$\pm$0.02\\
NGC 2976  & 103.5$\pm$13.6& 6.9$\pm$3.0  & 1.05$\pm$0.28\\
NGC 3198  & 158.1$\pm$0.7 & 16.4$\pm$0.3 & 0.82$\pm$0.03\\
NGC 3521  & 238.6$\pm$1.7 & 7.2$\pm$0.2  & 1.19$\pm$0.07\\
NGC 5055  & 209.1$\pm$0.6 & 8.5$\pm$0.2  & 0.36$\pm$0.01\\
NGC 6946  & 199.7$\pm$0.5 & 12.0$\pm$0.6 & 0.69$\pm$0.05\\
NGC 7331  & 257.8$\pm$1.7 & 3.5$\pm$1.3  & 0.12$\pm$0.04\\
\hline

\end{tabular}
\end{center}
\label{tab:rotcur}
\end{table}

We fit all the rotation curves of our sample galaxies with this Brandt profile. The black dashed lines in all the panels of Fig.~\ref{rotcur} shows the fit to the data. As it can be seen, for all the galaxies, the Brandt profile describes the rotation curves reasonably well. The fitted parameters are listed in Tab.~\ref{tab:rotcur}. To solve the hydrostatic equation we use these parametric values to represent the rotation curves.

\subsubsection{Dark matter halo parameters}

The next input parameter required to solve Eq.~\ref{eq_hydro} is the dark matter halo, which provides a considerable amount of gravity to the hydrostatic equation. For our sample galaxies, we adopt the mass models produced by \citet{deblok08}. \citet{deblok08} investigated the mass model of each galaxy extensively using both the NFW and the ISO profiles. However, none of these profiles could consistently describe all the galaxies better than the other one. Some galaxies found to be described better by an NFW profile whereas some galaxies are better suited for an ISO profile. Here, in our analysis, for any galaxy, we choose a model from \citet{deblok08} which describes the data better \citep[see Tab.3-6 of][for more details]{deblok08}. In Tab.~\ref{tab:dmpar} we present the chosen dark matter halo types and their parameters for our sample galaxies. In column (1) and (2) we list the name and the halo type we use respectively. Column (3) lists the central characteristic density of the dark matter halo whereas the characteristic radius is given in column (4). 

\begin{table}
\caption{Dark matter halo parameters}
\begin{center}
\begin{tabular}{lccc}

\hline
Name & DM halo & $R_c$ & $\rho_0$ \\
     &         & (kpc) & ($\times 10^{-3} $ \mspcc)\\
\hline
NGC 925   & ISO & 9.67  & 5.90\\
NGC 2841  & NFW & 20.55 & 12.40\\
NGC 2976  & ISO & 5.09  & 35.50\\
NGC 3198  & ISO & 2.71  & 47.50\\
NGC 3521  & ISO & 1.32  & 370.20\\
NGC 5055  & ISO & 11.73 & 4.80\\
NGC 6946  & ISO & 3.62  & 45.70\\
NGC 7331  & NFW & 60.20 & 1.05\\
\hline

\end{tabular}
\end{center}
\label{tab:dmpar}
\end{table}


\subsection{Solving the hydrostatic equation}
\label{sol_eqn}

With the inputs mentioned above, Eq.~\ref{eq_hydro} can be solved numerically. Eq.~\ref{eq_hydro} represents three second-order ordinary partial differential equations which are coupled through the first term in the RHS. We solve these equations simultaneously using 8$^{th}$ order Runge-Kutta method as implemented in python package {\tt scipy}. As these equations are second-order differential equations, one needs at least two initial conditions to solve it. The two conditions we employ here are  

\begin{equation}
\left( \rho_i \right)_{z = 0} = \rho_{i,0} \ \ \ \ {\rm and} \ \ \ \left(\frac{d \rho_i}{dz}\right)_{z=0} = 0
\label{init_cond}
\end{equation}

\noindent The second condition comes from the requirement of a static equilibrium, which demands no unbalanced force at the midplane and hence a density extrema at this point \citep[see, e.g.,][]{spitzer42}. On the other hand, the first condition in the above equation demands prior knowledge of the midplane density which is not known. This can be tackled using the knowledge of the observed surface density. 

We adopt a similar numerical procedure to solve the hydrostatic equation as used by many previous authors \citep{banerjee08,patra14,patra18a}. We iteratively solve the individual equations to get the correct midplane density $\rho_{i,0}$ which will produce the observed surface density. To solve individual disc components, for example, stars, we first assume a trial midplane density $\rho_{s,0}$ and solve Eq.~\ref{eq_hydro} to get a density solution $\rho_s(z)$. This solution is then integrated along $z$ to obtain the trial surface density, $\Sigma_s = 2 \int \rho_s \ dz$. This trial $\Sigma_s$ is then compared to the observed stellar surface density to update the next trail $\rho_{s,0}$. In this way, we iteratively approach to a $\rho_{s,0}$ such that it produces a $\Sigma_s$ which matches the observed value with 0.1\% accuracy. Using this approach, we found that the surface densities for all our sample galaxies at all radii converge within a few tens of iterations.

As Eq.~\ref{eq_hydro} represents three coupled equation, they should be solved in principle simultaneously. However, numerically it is not possible to solve it in a go. Instead, we solve the equation first for individual components and then introduce the solution of one component in the gravity term while solving the other components. For example, in the first iteration, we solve for the stars alone and obtain the density solution $\rho_{s,1}(z)$. While solving for this, we assume the density of \HI~and \hh~to be zero, i.e., no coupling of the gravitational force. Next, we solve for \HI~as an individual component. But this time, we no longer consider the density of the stars to be zero but we use $\rho_s(z) = \rho_{s,1}(z)$. However, for \hh, we still consider $\rho_{H2}(z)=0$. Thus we produce the solution for \HI, $\rho_{HI,1}(z)$ which have coupled the gravity of the stars only. Next, we solve for \hh~similarly considering the presence of both stars and \HI~by putting $\rho_s(z) = \rho_{s,1}(z)$ and $\rho_{HI}(z) = \rho_{HI,1}(z)$ in Eq.~\ref{eq_hydro} to obtain $\rho_{H2,1}(z)$. This marks the end of the first iteration. At the end of the first iteration, we have a slightly better coupled-solutions than expected otherwise for an individual case. In the second iteration onwards, solving for each disc component will have the coupling from the other components. We continue this iterative process until the solutions of every component converge to an accuracy better than 0.1\%. We note that for all our sample galaxies, the solutions quickly converges within a few iterations. 

Solving the hydrostatic equation is a computationally intensive step. It takes $\sim$ couple of minutes to solve Eq.~\ref{eq_hydro} in a standard workstation for a single radial point. However, as the vertical hydrostatic condition at any radius is not dependent on the same at any other radius, Eq.~\ref{eq_hydro} can be solved in parallel at different radii. Hence, we implement the solver using an MPI based parallel code for faster computation. 
 
\section{Results and discussion}

\begin{figure}
\begin{center}
\begin{tabular}{c}
\resizebox{0.45\textwidth}{!}{\includegraphics{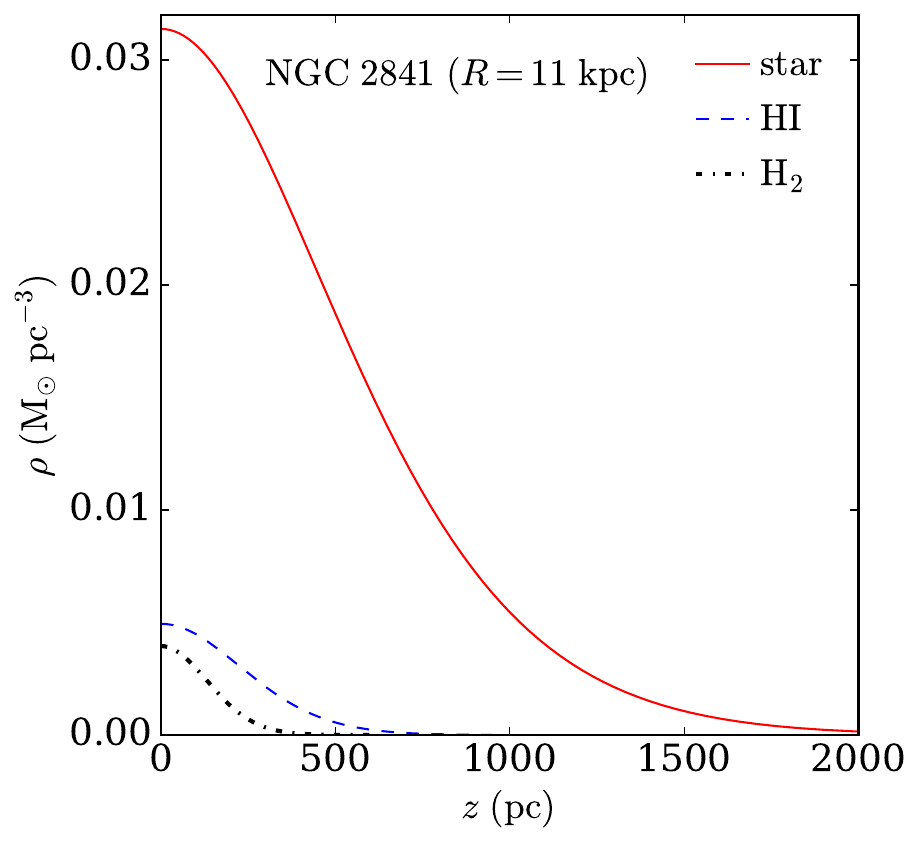}}
\end{tabular}
\end{center}
\caption{Example solutions of Eq.~\ref{eq_hydro} for the galaxy NGC 2841 at a radius of 11 kpc. The solid red line represents the mass density of stellar disc, whereas the blue dashed and the black dashed-dotted lines represent the atomic and the molecular mass densities respectively.}
\label{ex_soln}
\end{figure}

\begin{figure*}
\begin{center}
\begin{tabular}{c}
\resizebox{1.\textwidth}{!}{\includegraphics{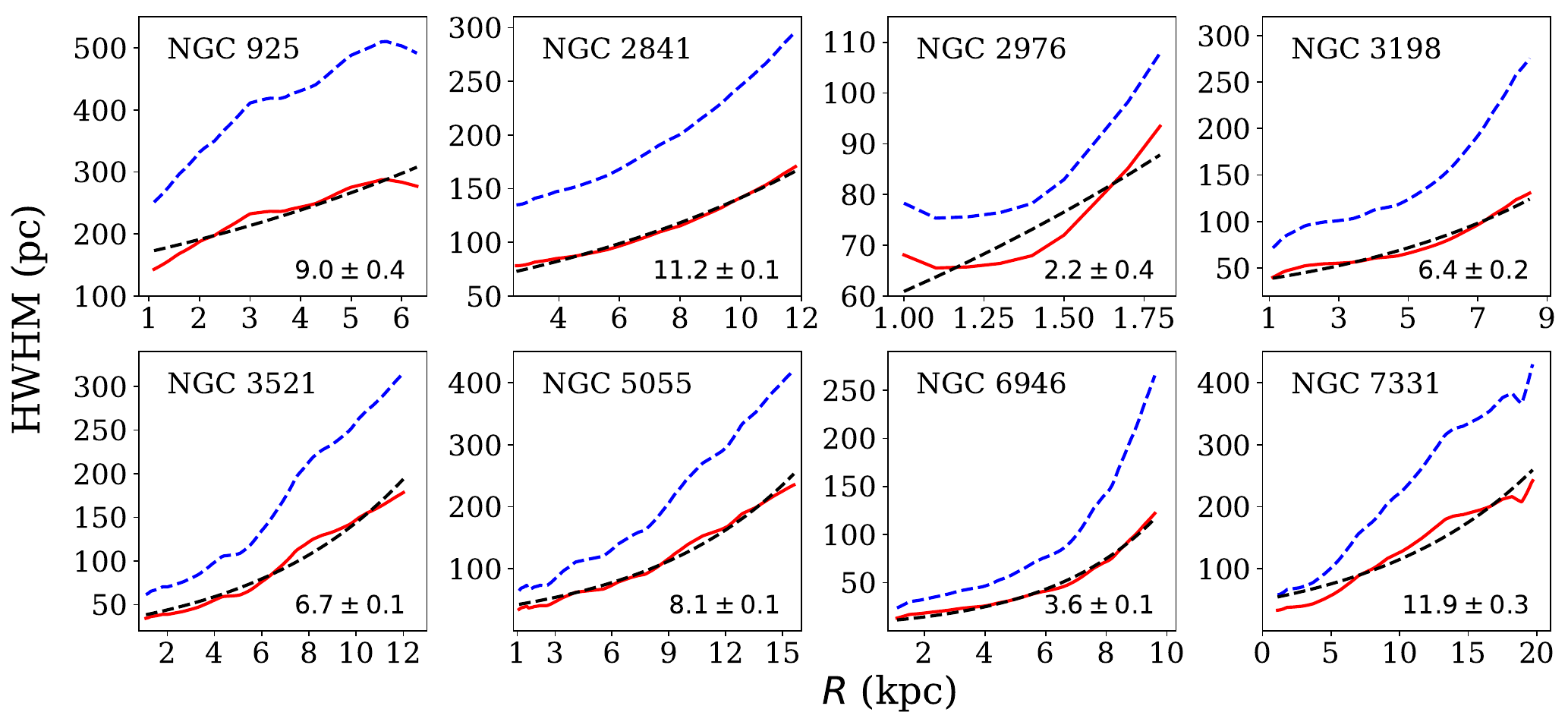}} 
\end{tabular}
\end{center}
\caption{The molecular and atomic scale heights of our sample galaxies. The blue dashed lines indicate the atomic scale heights whereas the solid red lines represent the molecular scale height. Each panel shows the scale heights for different sample galaxies as quoted at the top left corners of the respective panels. We fit the molecular scale heights with an exponential profile as shown by the dashed lines in the individual panels. The scale lengths of the fitted exponents are quoted in the bottom right corner of each panel in kpc.}
\label{sclh}
\end{figure*}

\begin{figure*}
\begin{center}
\begin{tabular}{c}
\resizebox{1.\textwidth}{!}{\includegraphics{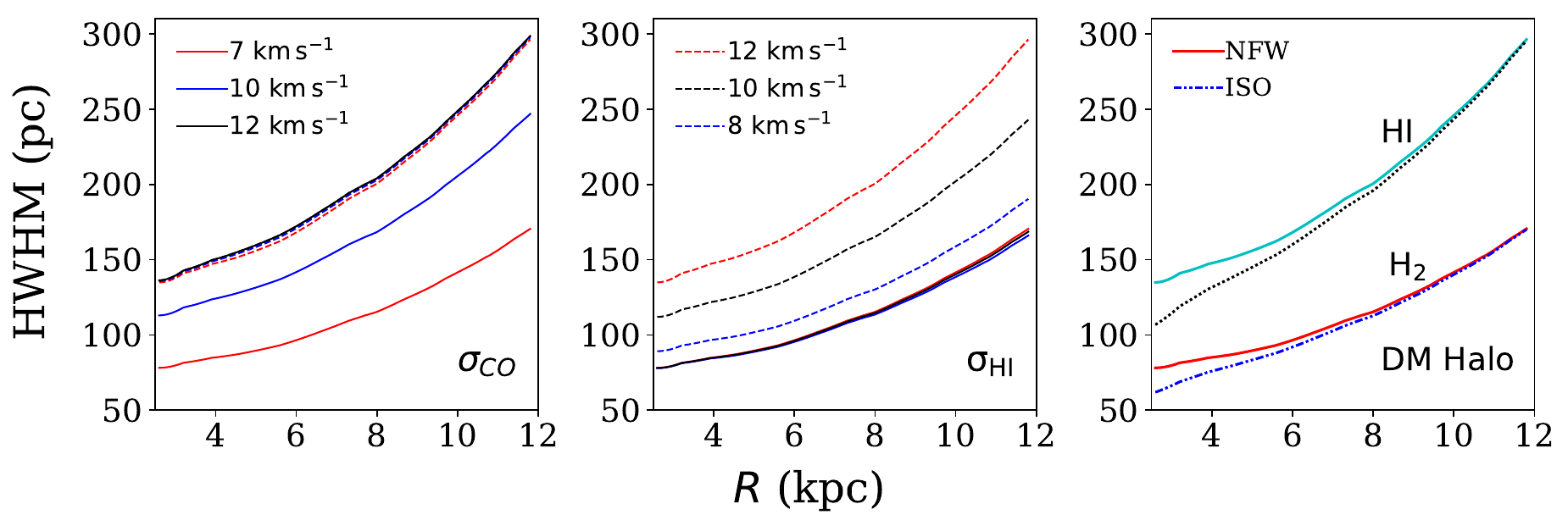}} 
\end{tabular}
\end{center}
\caption{The effect of assumed gas velocity dispersions and dark matter halo profiles on the molecular/atomic scale heights in NGC 2841. The left panel depicts the dependence of the gas scale heights on the assumed $\sigma_{CO}$ (for a constant $\sigma_{HI}=12$ \kms). The middle panel shows the dependence of the gs scale heights on $\sigma_{HI}$ (for a constant $\sigma_{CO}=7$ \kms). In both the panels the solid lines represent the molecular scale heights whereas the dashed lines indicate the atomic scale heights. Different colours represent differently assumed velocity dispersions as mentioned in the legends of the respective panels. As can be seen from these panels, the scale height of a particular gas disc considerably depends on its vertical velocity dispersion. However, it does not depend critically on the velocity dispersion of the other disc. In the right panel, we show the gas scale heights in NGC 2841 for different assumed dark matter halo profiles. The solid lines represent the estimated gas scale heights for an NFW halo (which is used in Fig.~\ref{sclh}, top row second panel) whereas the broken lines represent the same for an ISO halo. As can be seen, the gas scale heights moderately depend on the assumed dark matter halo profiles ($\lesssim 15-20\%$) only at the central regions. At outer radii, both the dark matter halo profiles produce very similar scale heights. See text for more details.}
\label{sclh_depend}
\end{figure*}

Using the method mentioned above we solve the hydrostatic equilibrium equation for our sample galaxies. In Fig.~\ref{ex_soln} we show an example solution of Eq.~\ref{eq_hydro} for the galaxy NGC 2841 at a radius of 11 kpc. As it can be seen from the figure, the stellar disc has a much higher thickness than the gaseous discs. The atomic disc is found to be thicker than the molecular disc as well (this is true for all radii). We note that for a single component isothermal disc, the solutions follow a $sech^2$ law \citep[see][for detailed calculations]{bahcall84a,bahcall84b}, however, in the presence of coupling, the solution deviates from a $sech^2$ law and tend to follow a Gaussian-like profile. 

We note that the assumption of the hydrostatic equilibrium is crucial to our analysis and any violation of this assumption would lead to a wrong interpretation of the results. Observationally it has been found that the highly energetic activities with observed outflows and enhanced star formations are generally confined well within the central 1 kpc region of a spiral galaxy \citep{bolatto13,irwin96}. Within this region, the assumption of a hydrostatic equilibrium might not hold good. Hence, while solving Eq.~\ref{eq_hydro} for our sample galaxies, we exclude a region of central 1 kpc. It should be noted that at places in small regions, the hydrostatic condition might not hold good due to various reasons, but, as we are using azimuthally averaged quantities, it is expected that these small fluctuations would have a negligible effect on the global stability. For NGC 2841, CO was not detected within a central region of radius 2.6 kpc and hence, for this galaxy, we only solve Eq.~\ref{eq_hydro} outside this region. 

As the solutions of the hydrostatic equation (i.e., $\rho_s, \ \rho_{HI}$ and $\rho_{H_2}$ as a function of $R$ and $z$) provide a detailed three-dimensional distribution of the different disc components, it can be used to estimate the atomic and molecular scale heights in our sample galaxies. The scale height is defined as the Half Width at Half Maxima (HWHM) of the vertical density distribution in the baryonic discs. This scale height is an indicative measure of the thickness of the atomic or the molecular disc. Using the solutions for the atomic and the molecular gas discs for our sample galaxies, we calculate the HWHM of the vertical density distribution as a function of radius. Thus estimated atomic and molecular scale heights are plotted in Fig.~\ref{sclh}. It can be seen from the figure that the scale heights of the molecular discs vary between 20-100 pc in the central region ($R \sim$ few kpc) and $\sim$ 150-200 pc at the outer regions ($R \sim 10-15$ kpc). The atomic scale heights are found to be higher than the molecular scale height by a factor of $\sim$ 2 at all radii. The molecular scale height found for our sample galaxies are consistent with that of the Milky Way \citep[see, e.g.,][]{nakanishi06}. In spite of being having very different surface density profiles and dark matter distribution, the molecular scale heights of our sample galaxies show a very similar trend. For example, for almost all the galaxies, molecular scale heights start $\sim$ 50 pc at the central region and increases to $\sim$ 150-200 pc at the outer regions.

\begin{figure}
\begin{center}
\begin{tabular}{c}
\resizebox{.45\textwidth}{!}{\includegraphics{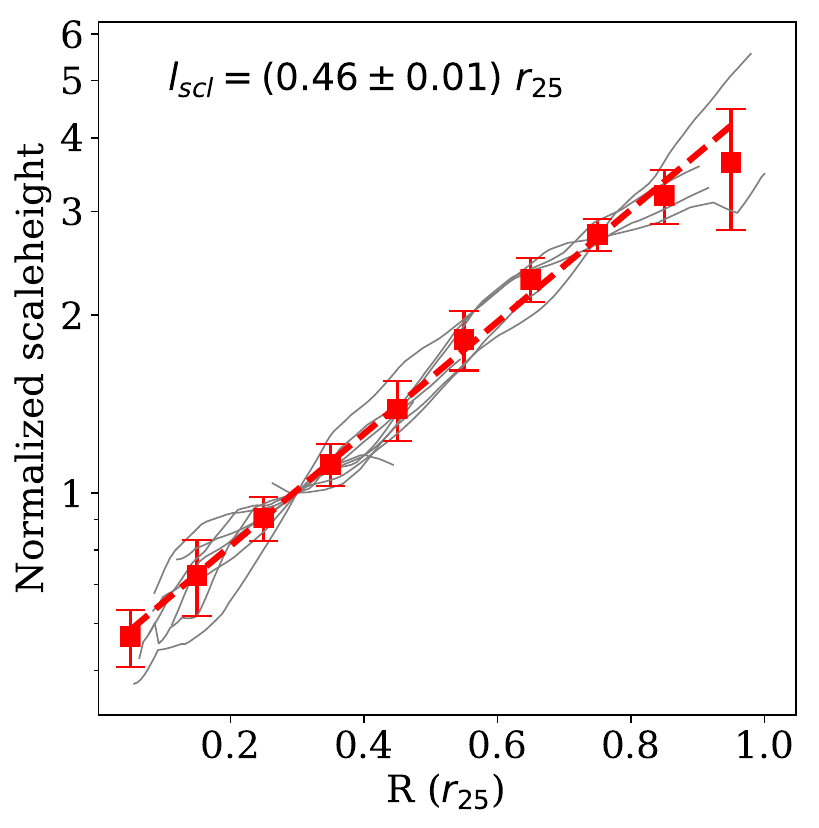}} 
\end{tabular}
\end{center}
\caption{Normalised scale height (HWHM) profiles for our sample galaxies. The solid lines represent the HWHM profiles for our sample galaxies normalised to the value 1 at a radius of 0.3 $r_{25}$, the 25th magnitude B-band isophote. The red squares with error bars represent the mean normalised scale height at radial bins of 0.1 $r_{25}$. The error bars represent 1-$\sigma$ scatter on the data. The red dashed line represents an exponential fit to the mean normalised scale height. We find that the scale height follows a tight exponential function with a scale length of $0.46 \pm 0.01 \ r_{25}$.}
\label{sclh_norm}
\end{figure}

\begin{figure}
\begin{center}
\begin{tabular}{c}
\resizebox{.47\textwidth}{!}{\includegraphics{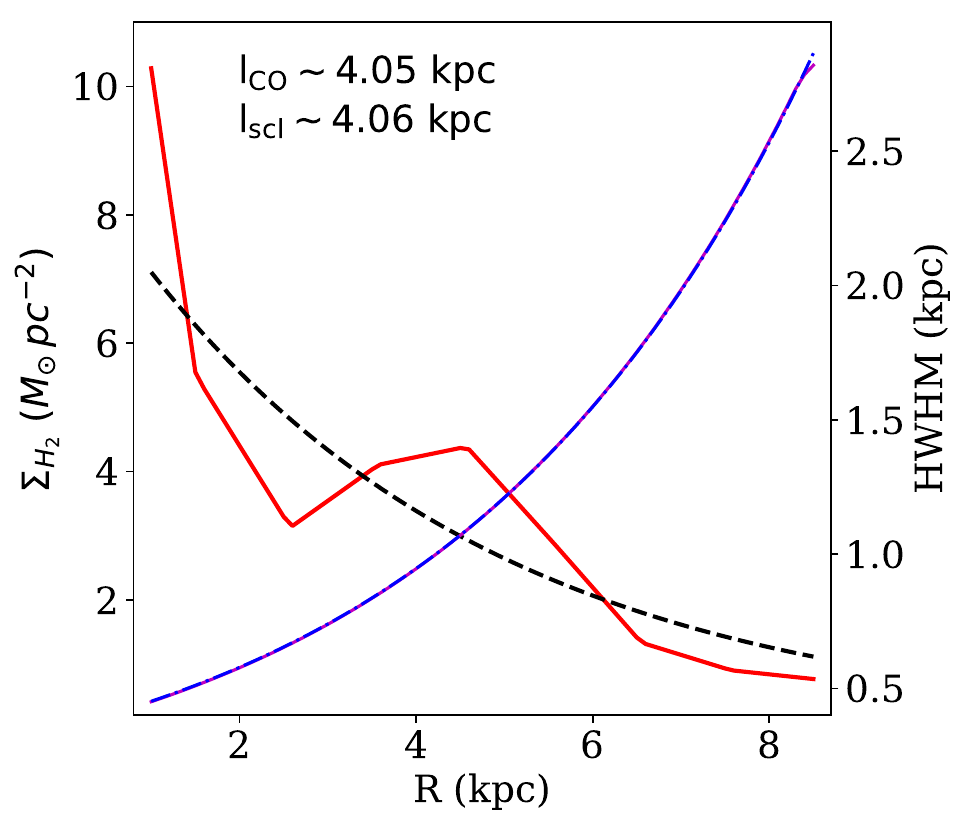}} 
\end{tabular}
\end{center}
\caption{The behaviour of the molecular scale height in the absence of any external influence in NGC 3198. The solid red and the black dashed lines represent the molecular gas surface density profile ($\Sigma_{H_2}$), and it's fit to an exponential profile respectively (marked by the left axis). The solid magenta and the dashed-dotted blue lines represent the estimated molecular scale height (HWHM) and its fit to a rising exponential function respectively (marked by the right axis). The scale lengths of the exponential fits to the $\Sigma_{H_2}$ ($l_{CO}$) and the HWHM profile ($l_{scl}$) are found to be $\sim 4.05$ kpc and $\sim 4.06$ kpc (as quoted in the top left), are almost the same.}
\label{sclh_test}
\end{figure}

\begin{figure*}
\begin{center}
\begin{tabular}{ccc}
\resizebox{0.34\textwidth}{!}{\includegraphics{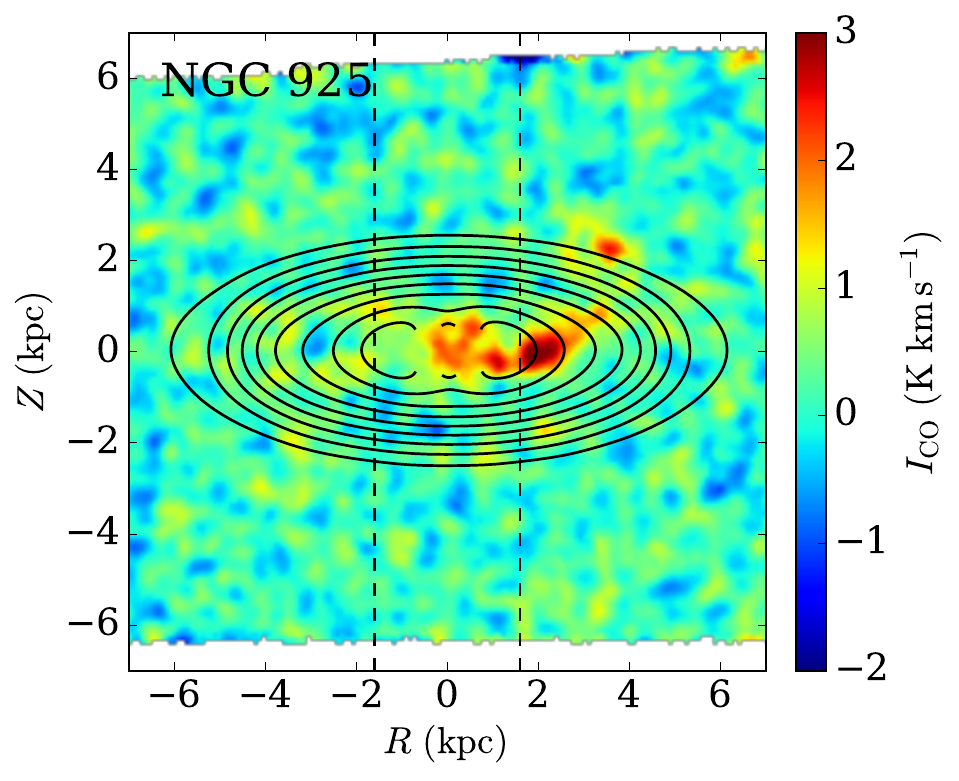}} & 
\resizebox{0.34\textwidth}{!}{\includegraphics{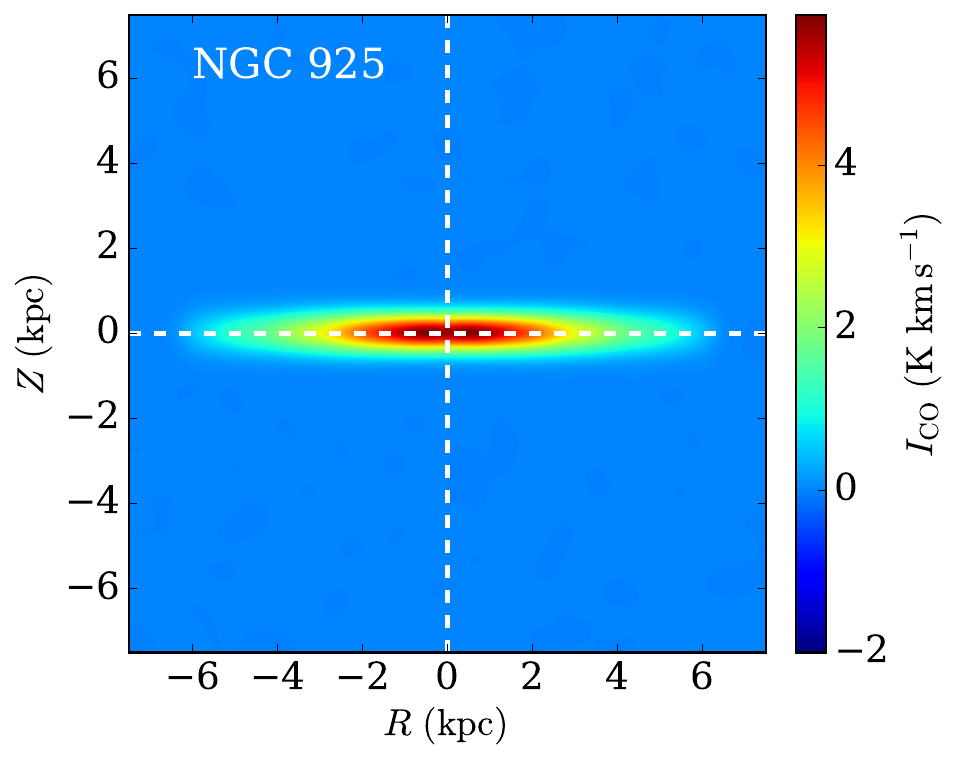}} &
\resizebox{0.27\textwidth}{!}{\includegraphics{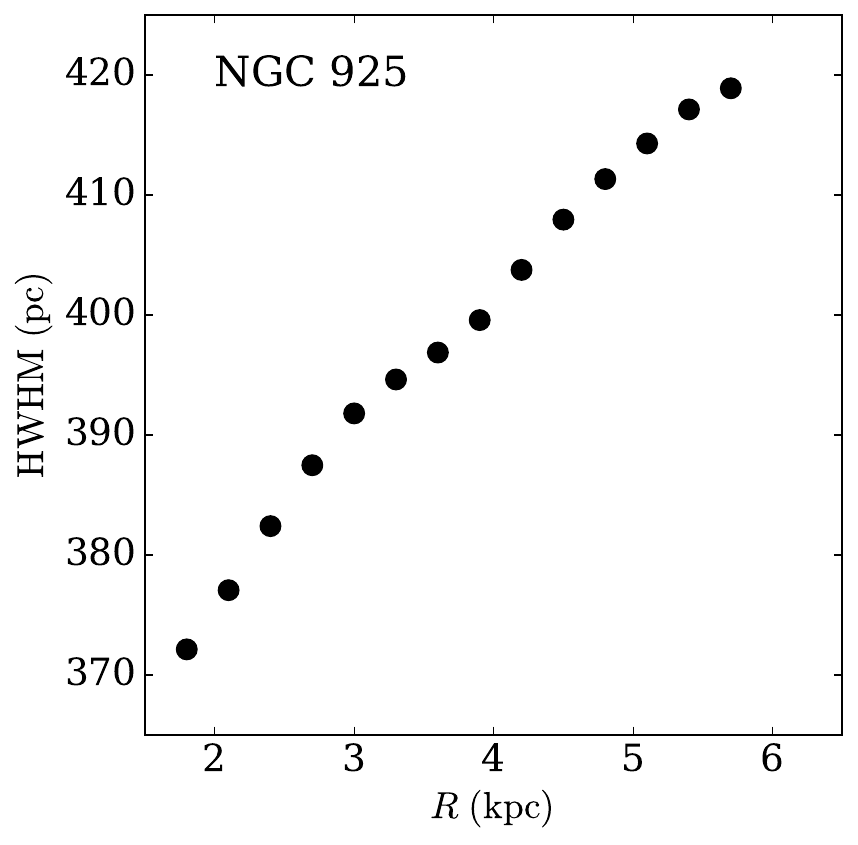}} \\

\resizebox{0.34\textwidth}{!}{\includegraphics{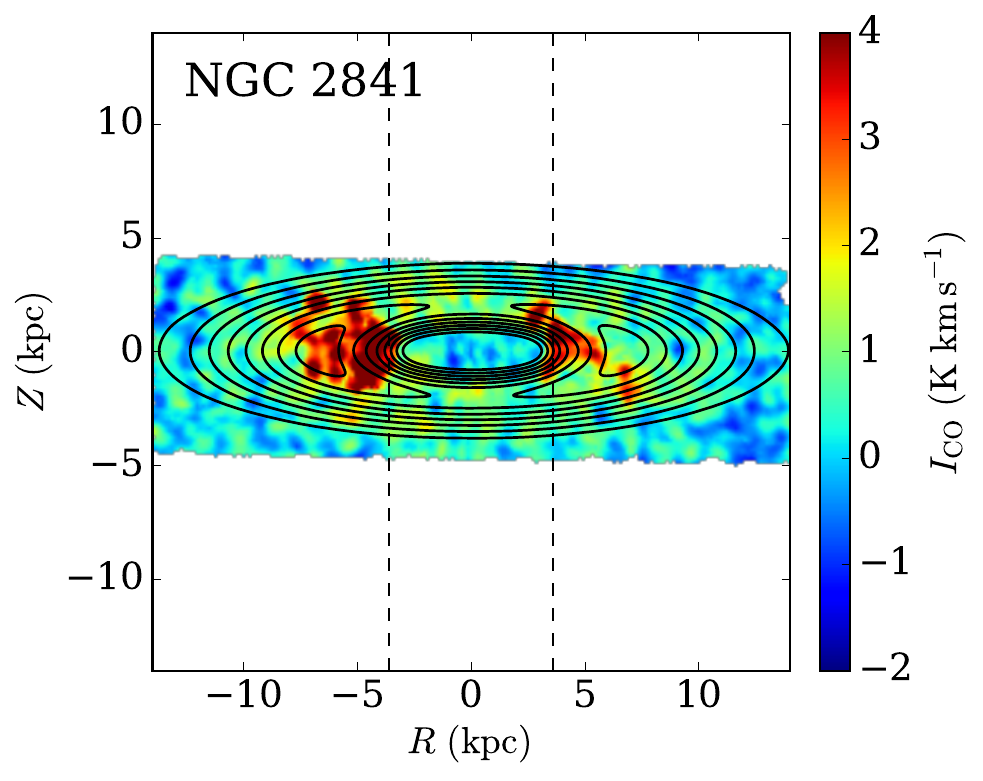}} & 
\resizebox{0.34\textwidth}{!}{\includegraphics{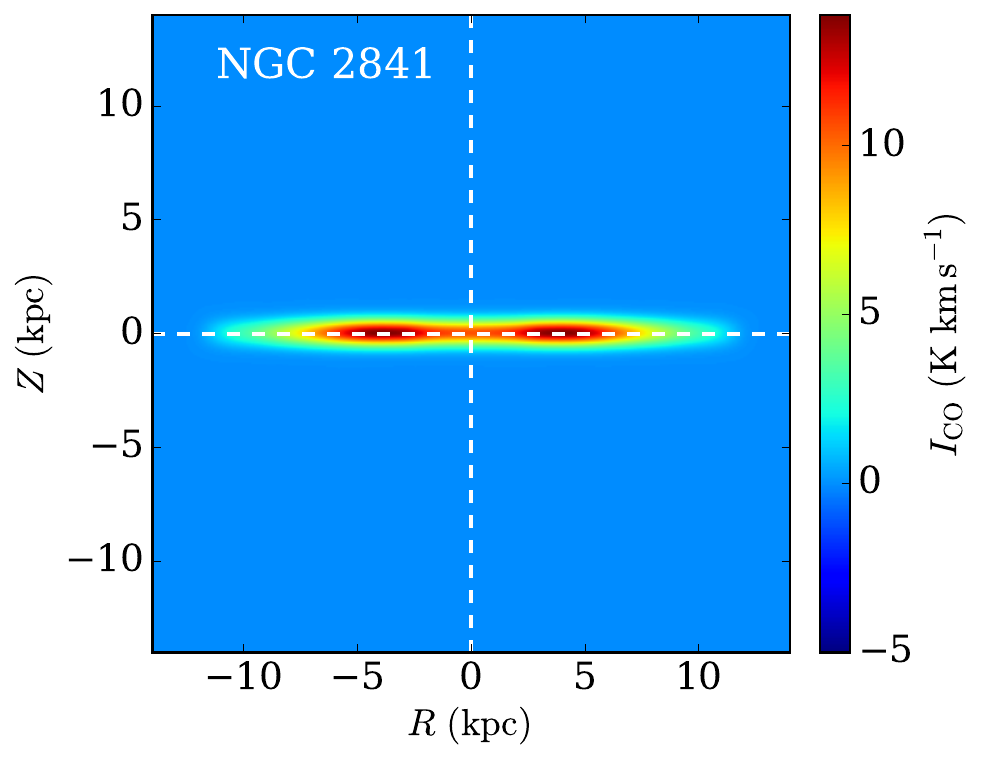}} &
\resizebox{0.27\textwidth}{!}{\includegraphics{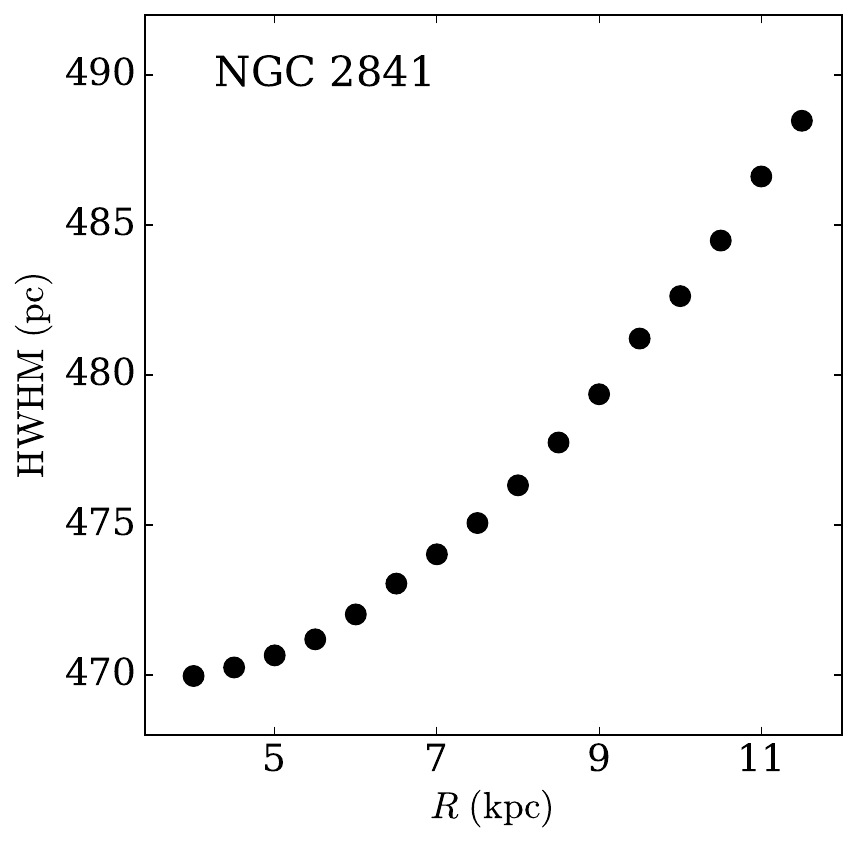}} \\

\resizebox{0.34\textwidth}{!}{\includegraphics{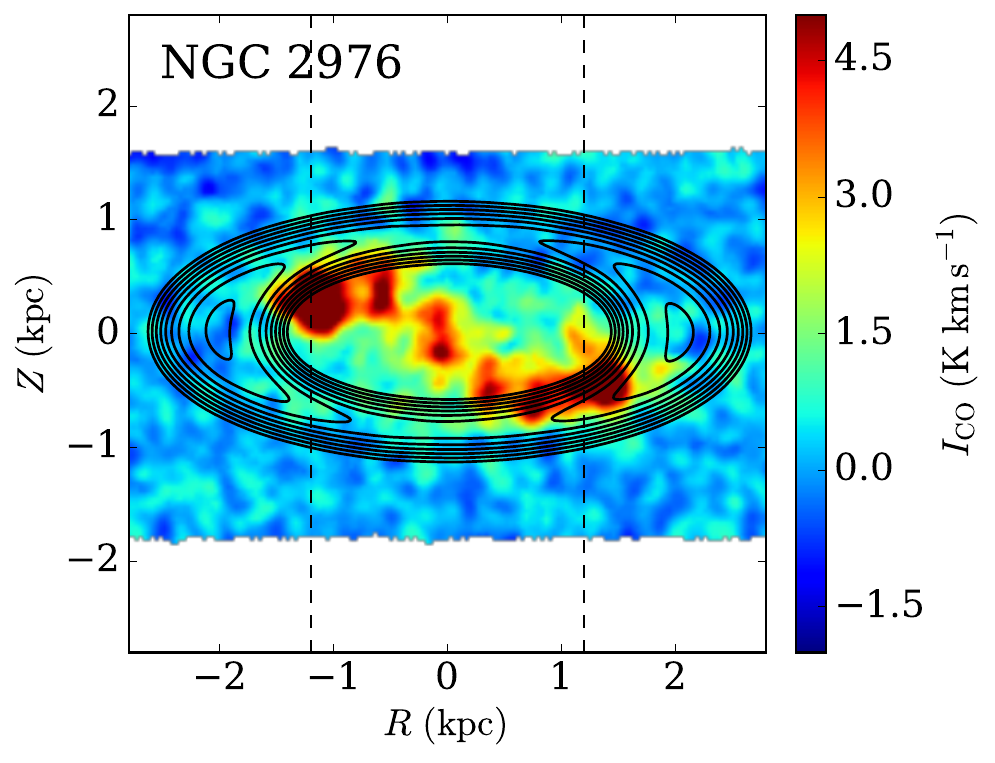}} & 
\resizebox{0.34\textwidth}{!}{\includegraphics{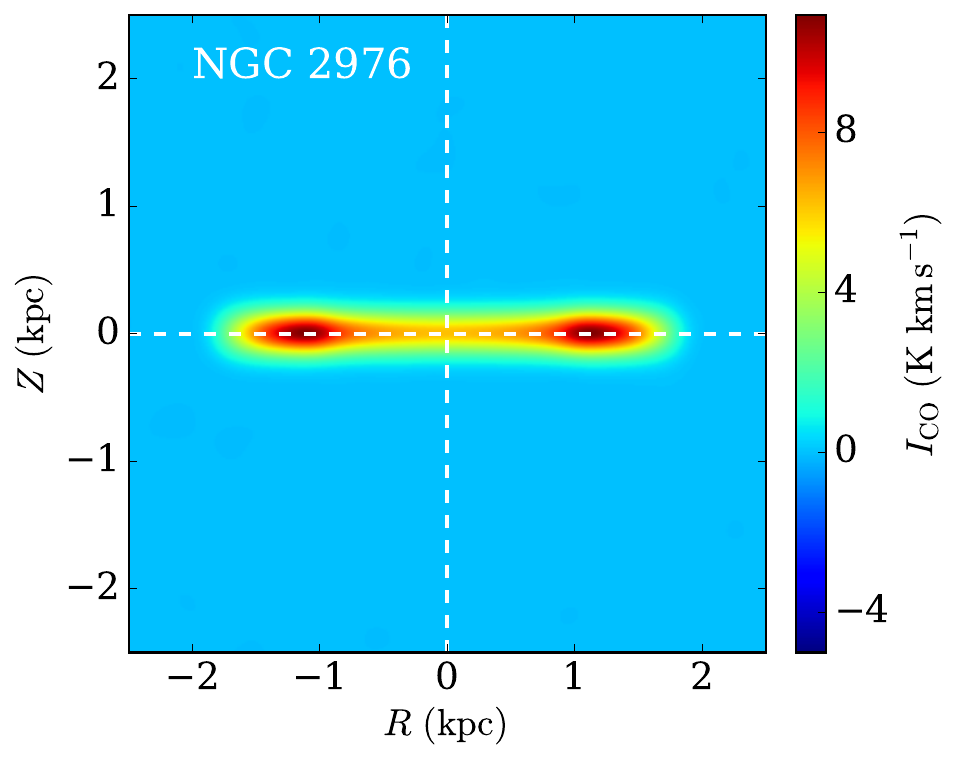}} &
\resizebox{0.27\textwidth}{!}{\includegraphics{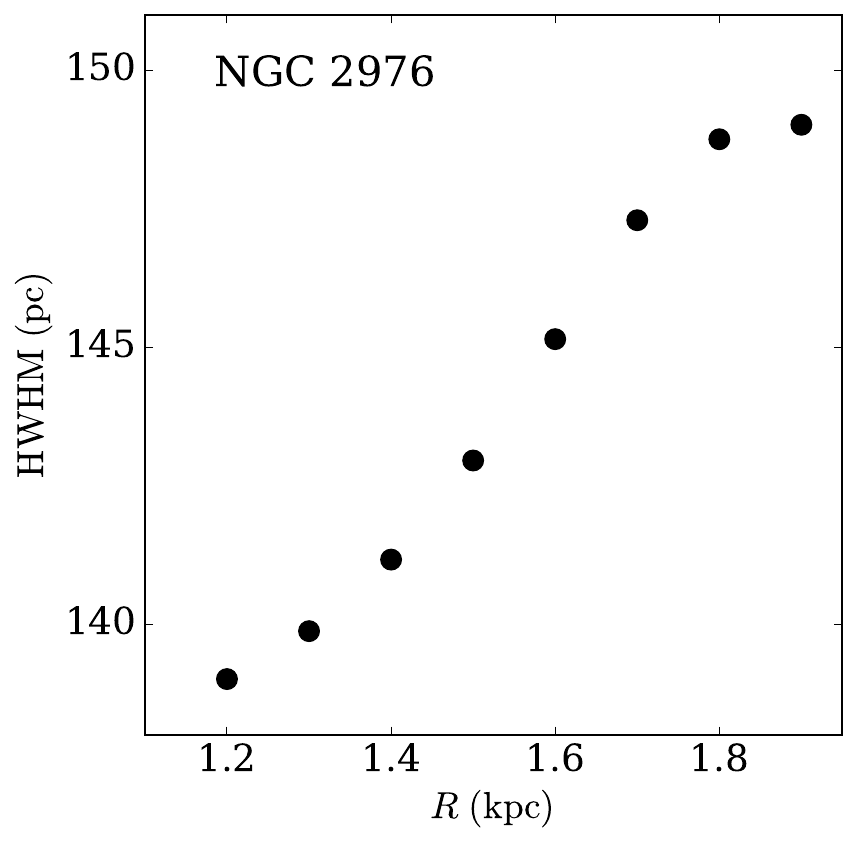}} \\

\resizebox{0.34\textwidth}{!}{\includegraphics{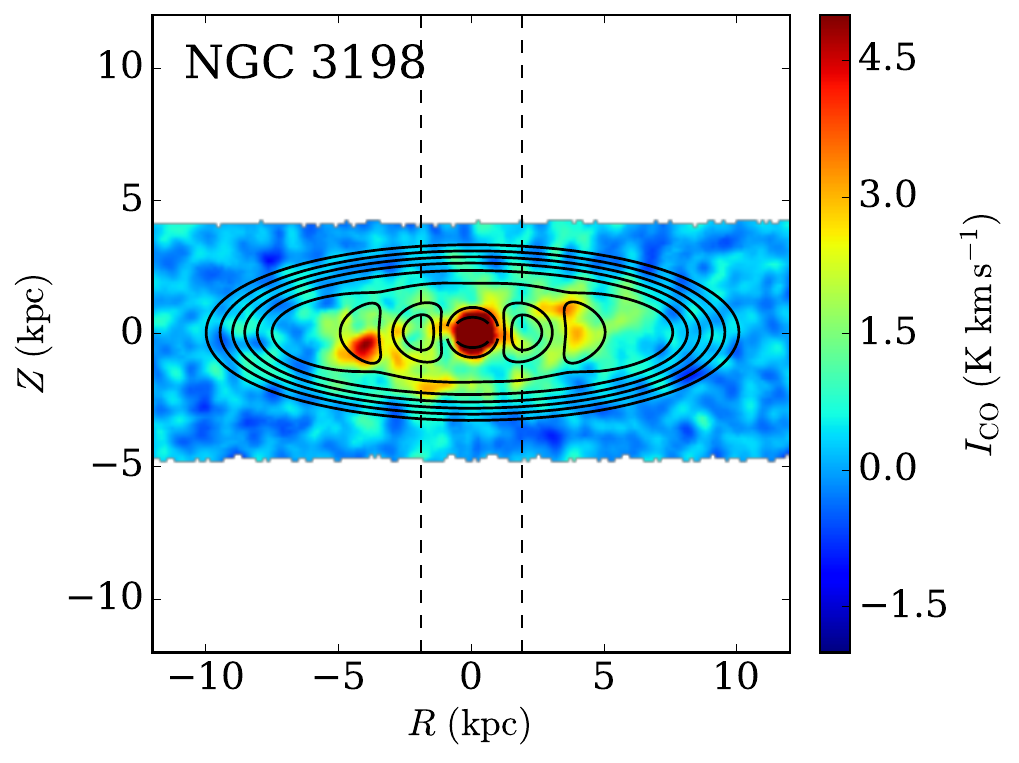}} & 
\resizebox{0.34\textwidth}{!}{\includegraphics{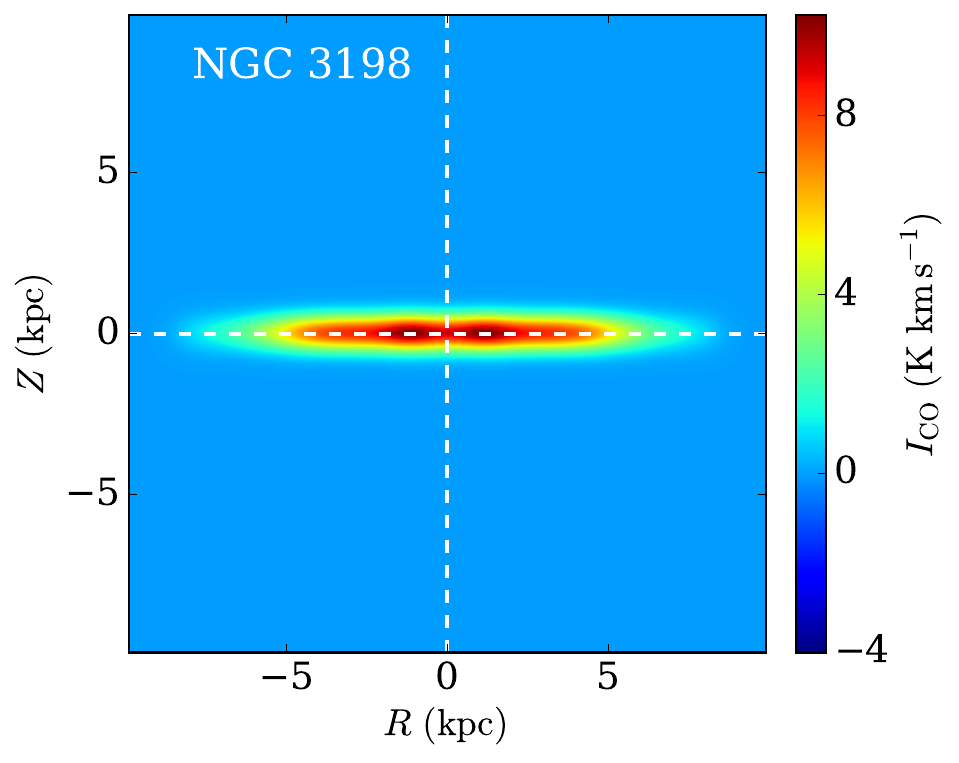}} &
\resizebox{0.27\textwidth}{!}{\includegraphics{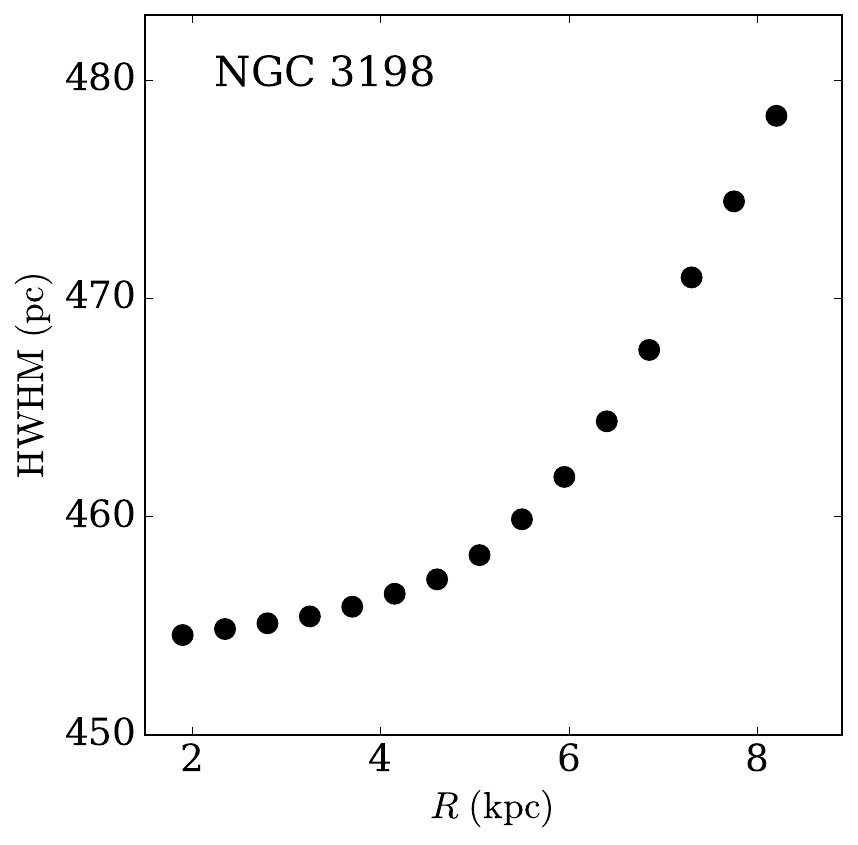}} \\

\resizebox{0.34\textwidth}{!}{\includegraphics{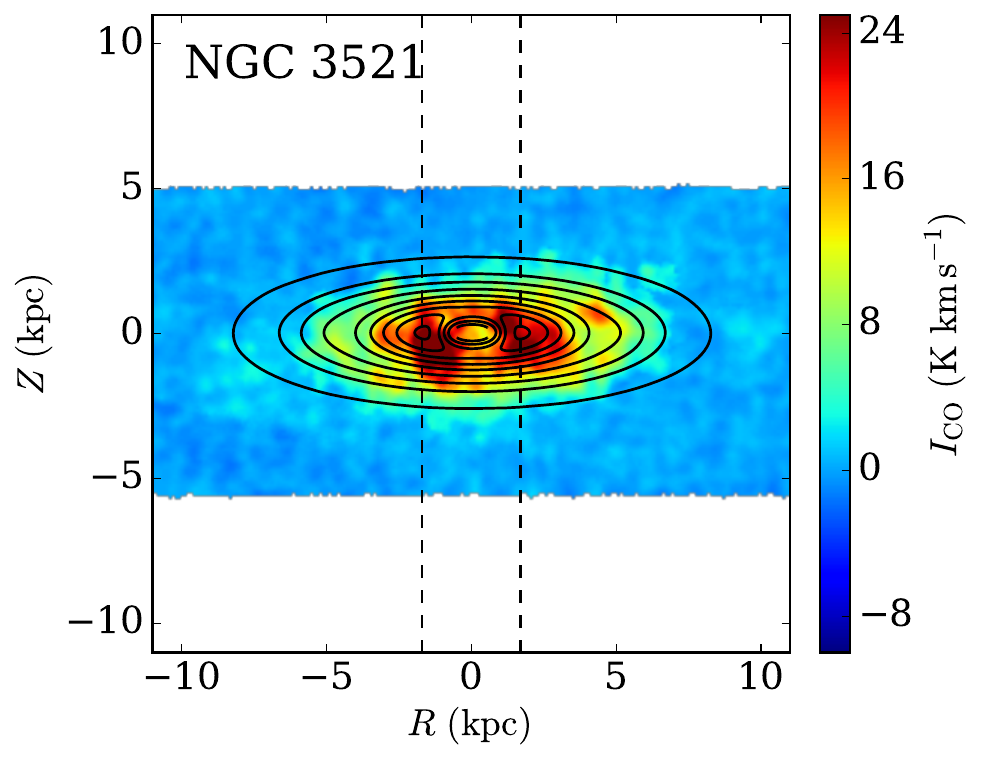}} & 
\resizebox{0.34\textwidth}{!}{\includegraphics{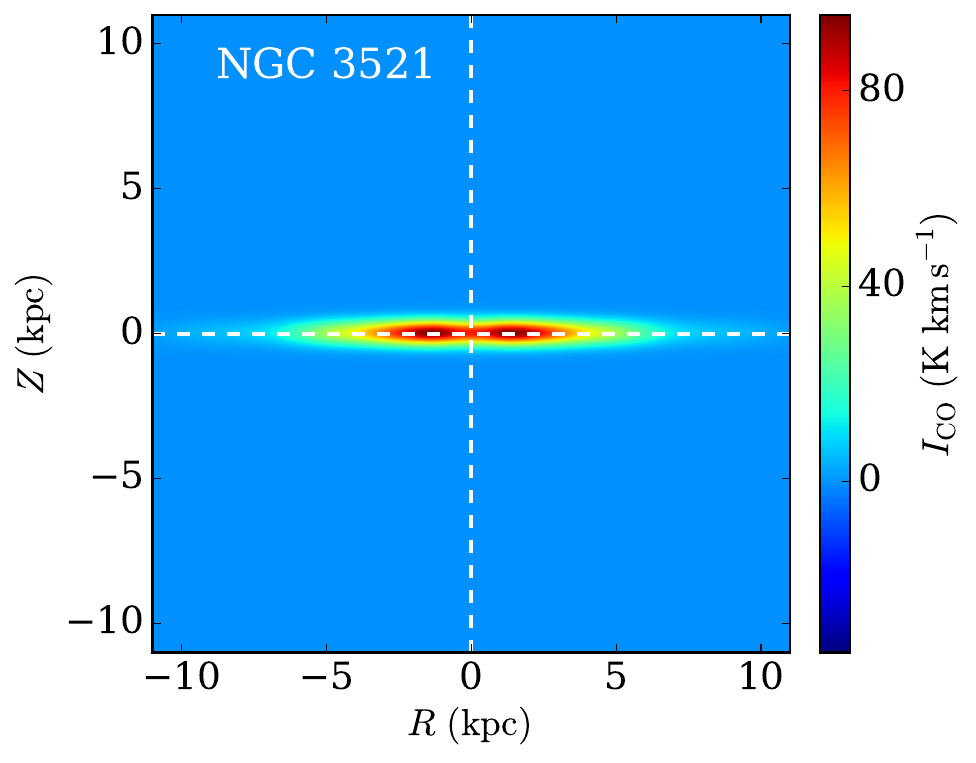}} &
\resizebox{0.27\textwidth}{!}{\includegraphics{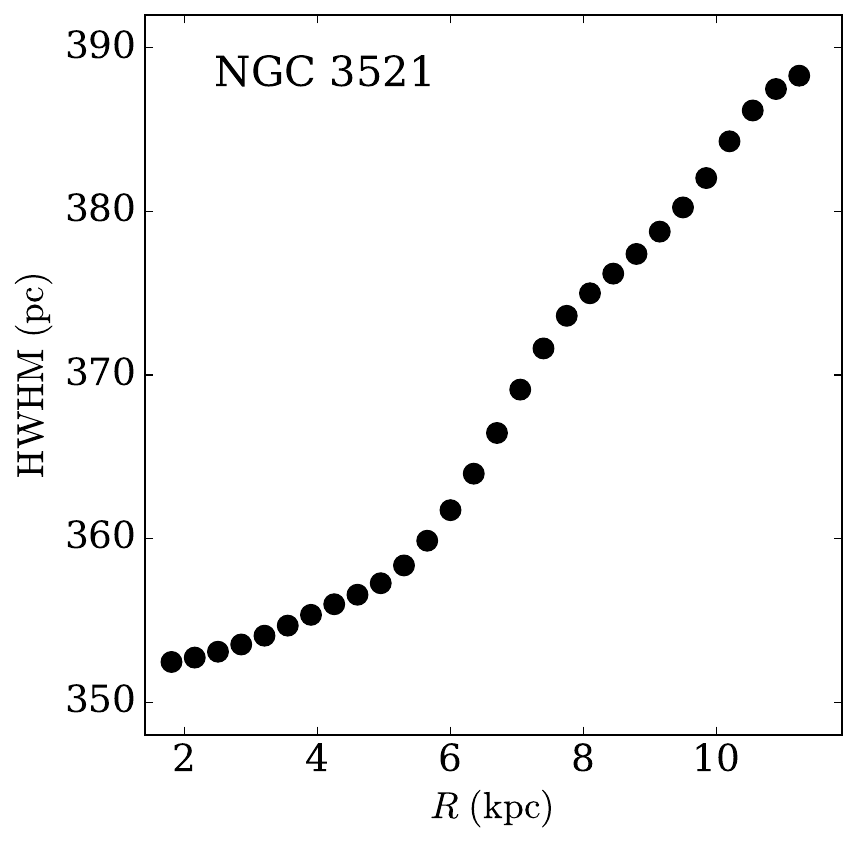}} \\

\end{tabular}
\end{center}
\caption{Left column: Comparison of the modelled column density maps with the observed ones. Black contours represent the modelled molecular discs whereas the colour scales represent the molecular discs as observed in the HERACLE survey. The contour levels are NGC 925: (0.7, 0.8, 0.9, ...), NGC 2841: (1., 1.3, 1.6, ...), NGC 2976: (1.2, 1.4, 1.6, ...), NGC 3198: (1., 1.25, 1.5, ...), NGC 3521: (1.5, 4.7, 7.9, ...) in the units of $K \ km \thinspace s^{-1}$. Middle column: Shows the molecular column density maps of our sample galaxies when viewed edge-on. Right column: Shows the HWHM profiles of the edge-on column density distribution as a function of radius.}
\label{ovrplt1}
\end{figure*}

\begin{figure*}
\begin{center}
\begin{tabular}{ccc}
\resizebox{0.34\textwidth}{!}{\includegraphics{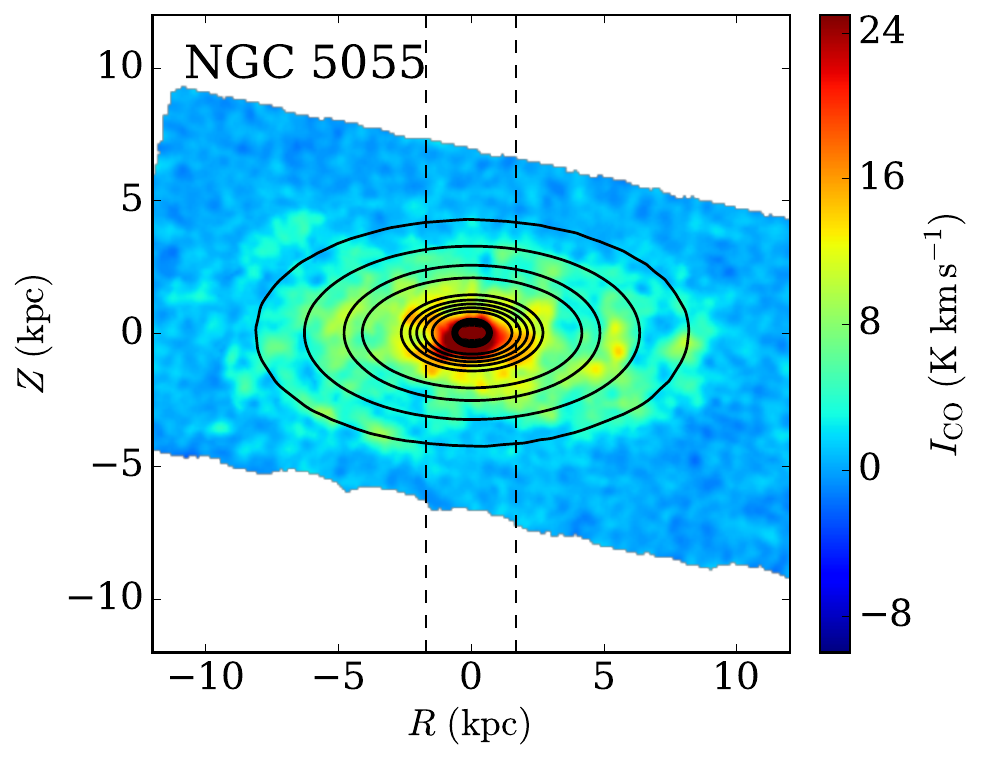}} & 
\resizebox{0.34\textwidth}{!}{\includegraphics{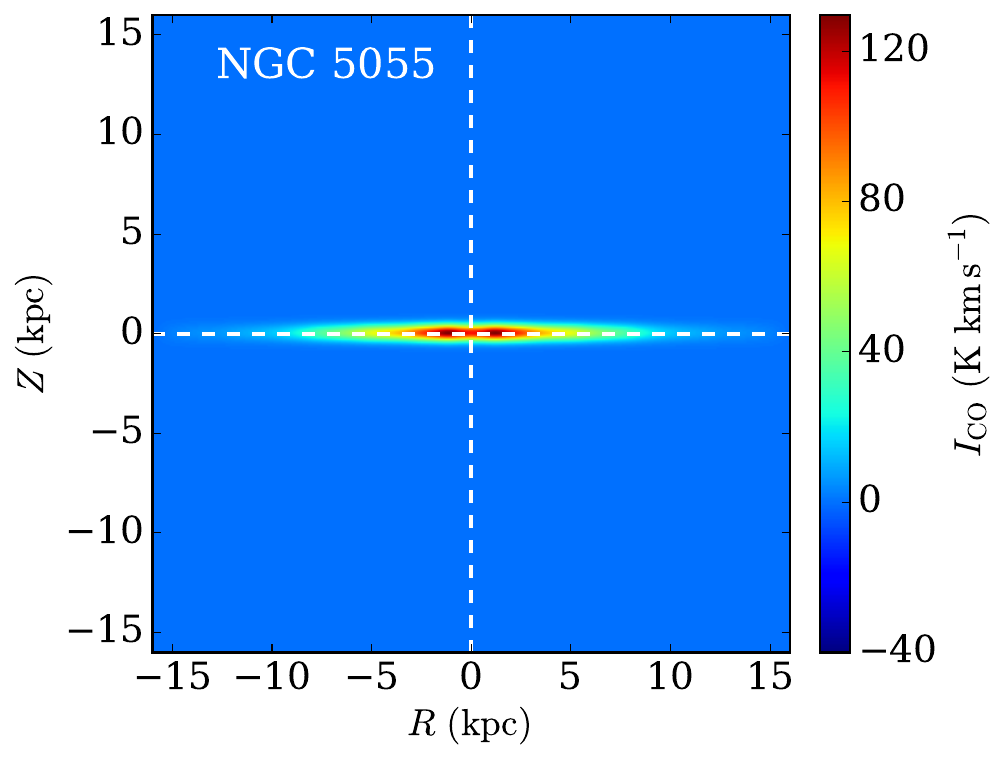}} &
\resizebox{0.27\textwidth}{!}{\includegraphics{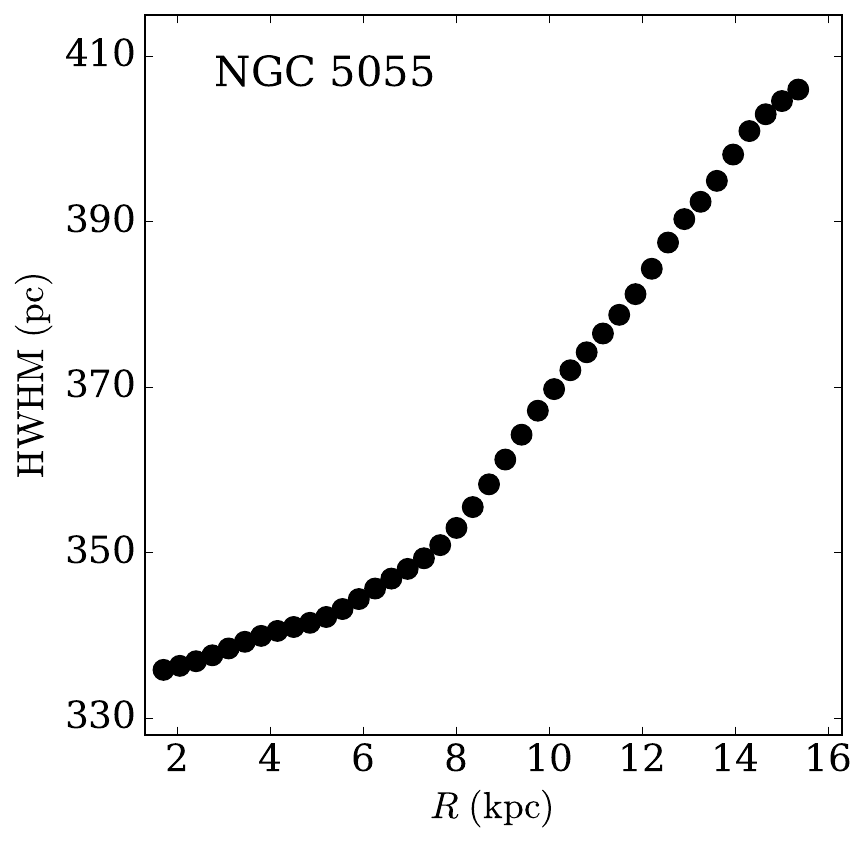}} \\

\resizebox{0.34\textwidth}{!}{\includegraphics{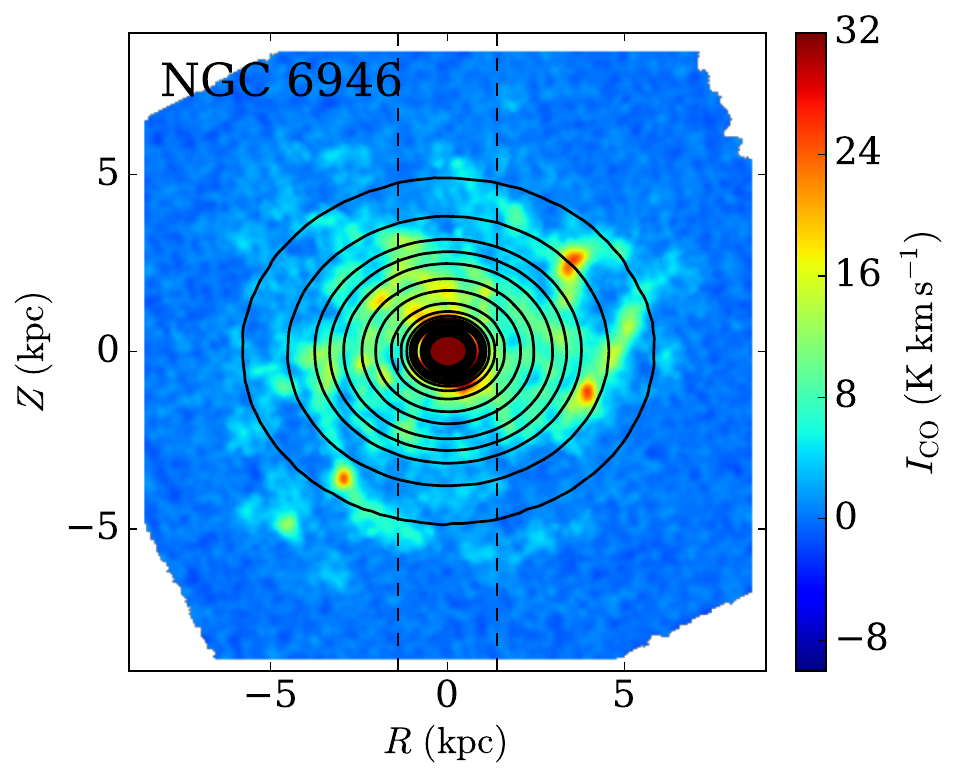}} & 
\resizebox{0.34\textwidth}{!}{\includegraphics{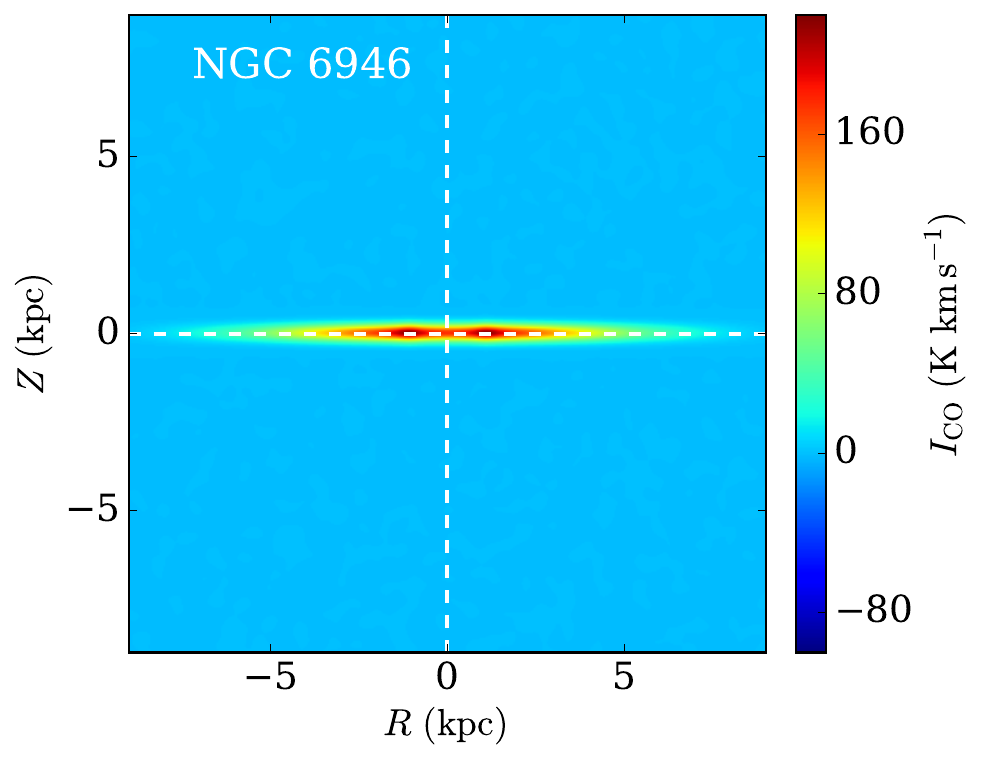}} &
\resizebox{0.27\textwidth}{!}{\includegraphics{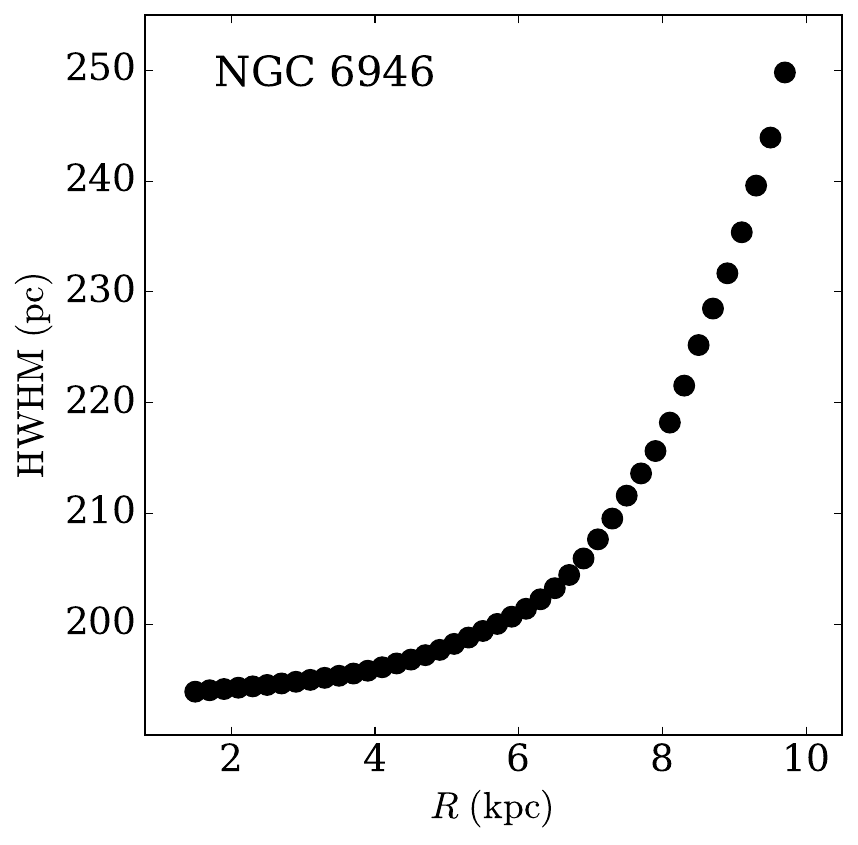}} \\

\resizebox{0.34\textwidth}{!}{\includegraphics{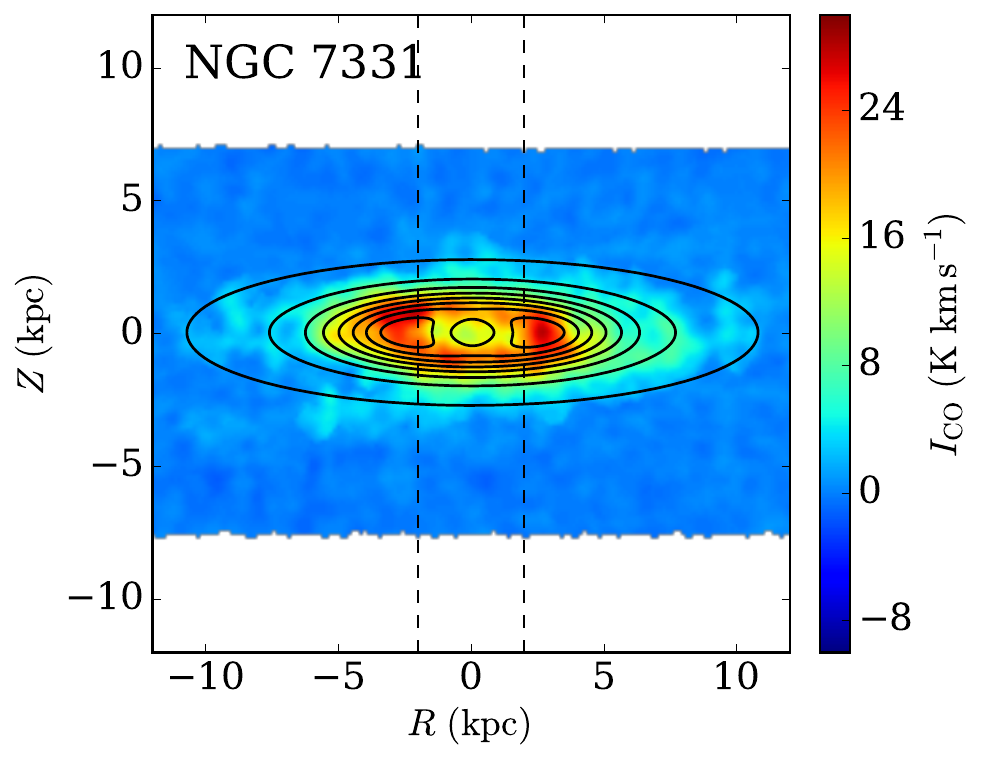}} & 
\resizebox{0.34\textwidth}{!}{\includegraphics{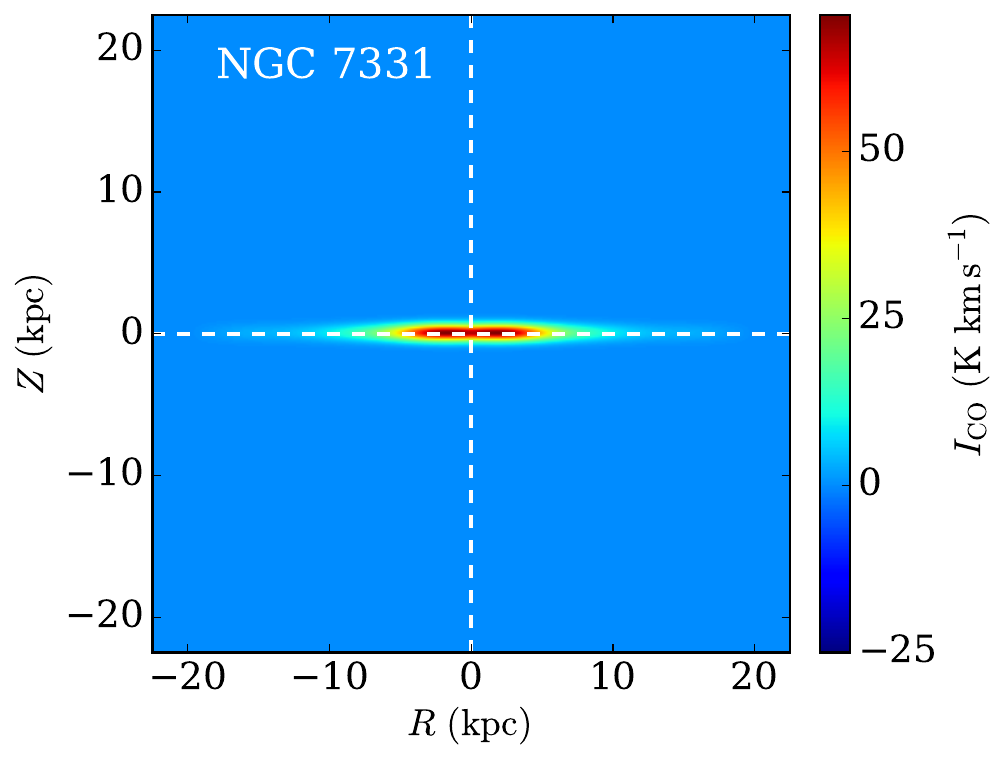}} &
\resizebox{0.27\textwidth}{!}{\includegraphics{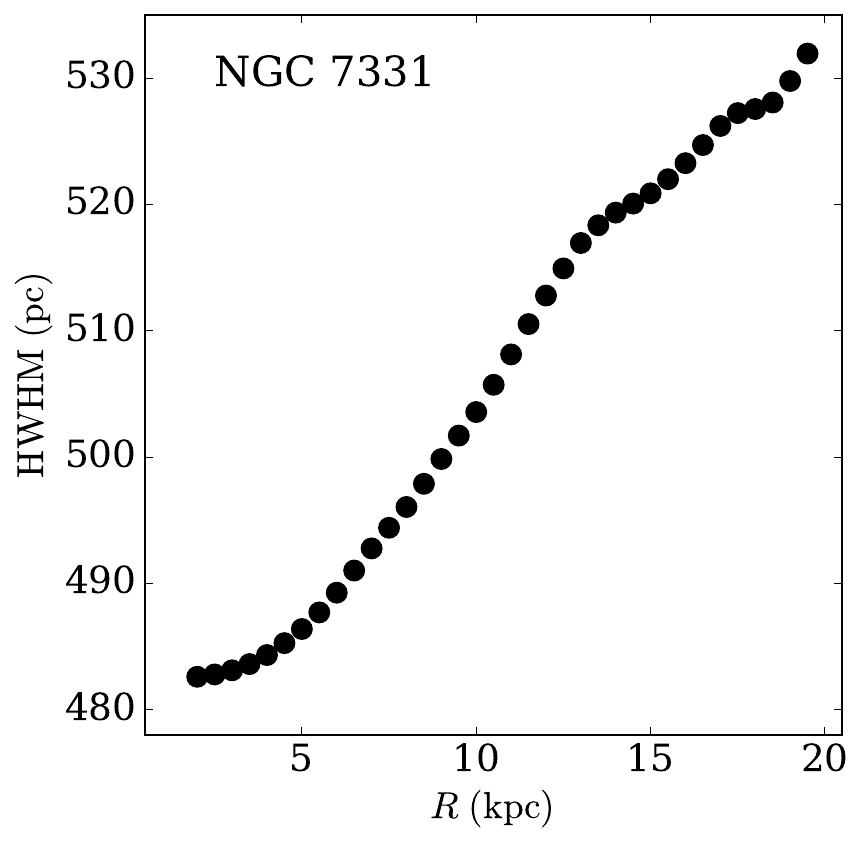}} \\

\end{tabular}
\end{center}
\caption{Same as Fig.~\ref{ovrplt1}. The contour levels in the left column are NGC 5055: (1., 2.9, 4.8, ...), NGC 6946: (0.5, 1.8, 3.0, ...) in the units of $K \ km \thinspace s^{-1}$.}
\label{ovrplt2}
\end{figure*}

As mentioned earlier in \S3.2.2, constant gas velocity dispersions of $\sigma_{CO} = 7$ \kms~and $\sigma_{HI} = 12$ \kms are used to solve the hydrostatic equilibrium equation. However, the gas velocity dispersions are found to vary from galaxy to galaxy and within a galaxy \citep{tamburro09,calduprimo13,mogotsi16} which might influence the estimated scale heights considerably. To investigate this in more details, we solve Eq.~\ref{eq_hydro} for a representative galaxy, NGC 2841 assuming different atomic and molecular velocity dispersions and estimate the scale heights. In Fig.~\ref{sclh_depend} we show the dependence of the gas scale heights on the assumed velocity dispersions. The left panel depicts the variation of the atomic and the molecular scale heights for a set of assumed $\sigma_{CO}$. Whereas, the middle panel shows the same but for a set of assumed $\sigma_{HI}$. It is found that $\sigma_{CO}$ significantly influences the molecular scale height. A few \kms~change in the $\sigma_{CO}$ can introduce a $\sim 40-50\%$ change in the molecular scale height. However, its effect on the atomic scale height is negligible. Similarly, the $\sigma_{HI}$ is found to influence the atomic scale height considerably though it does not affect the molecular scale height meaningfully. 

As for our sample galaxies, both the NFW and the ISO dark matter halos can explain the observed rotation curve reasonably within the measurement uncertainties; no particular profile is observationally much preferred to describe the dark matter halos in these galaxies over the other. Hence, it is worth exploring the effect of the chosen dark matter halo profile on the estimated gas scale heights. To do that, we solve Eq.~\ref{eq_hydro} for NGC 2841 considering both the NFW and the ISO dark matter halo profiles (from Table. 5 of \citet{deblok08}) and estimate the scale heights. In the right panel of Fig.~\ref{sclh_depend} we plot the resulting atomic and molecular scale heights. As it can be seen from the figure, a different dark matter halo profile (ISO) for NGC 2841 produces a maximum difference of $\sim 15-20\%$ in the molecular scale height at inner radii. This difference gradually narrows down as one moves towards outer radii. As the structural difference between the dark matter halos is maximum at the centre (core or cusp), the scale heights due to these two halos differ the most in the central region.

Next, to examine the nature of the flaring in the molecular discs of our sample galaxies, we fit the scale heights with an exponential profile of the form $h_{scl}(R) = h_0 \exp(R/R_0)$. Where $h_0$ is the characteristic scale height and $R_0$ is the exponential scale length. In each panel of Fig.~\ref{sclh}, the exponential fits to the scale heights are shown by the black dashed lines. As can be seen from the figure, for all our sample galaxies an exponential profile well describe the molecular scale height except for NGC 2976. For this galaxy, we only solve the hydrostatic equation from 1 kpc to 1.75 kpc which is a much smaller region as compared to our other sample galaxies and might not be good enough to capture the full variation of a larger region. The scale lengths of the exponential fits ($R_0$) are quoted at the bottom right corners of the respective panels in the units of kpc.

To investigate how universal is this exponential flaring within our sample galaxies, we normalize all the scale heights for our galaxies and plot as a function of radius. Adopting a similar approach as used by \citet{schruba11}, we normalize all the scale heights to have a value 1 at a radius of $0.3 \ r_{25}$, where $r_{25}$ is the optical radius of a galaxy calculated by fitting the 25th magnitude B-band isophote. In Fig.~\ref{sclh_norm} we plot these normalized scale heights (solid lines) as a function of radius (in the units of $r_{25}$). It is very interesting to see from the figure that, all our sample galaxies follow a fairly tight exponential law when plotted against radius in an absolute unit of $r_{25}$. To obtain a universal law, we calculate the mean scale heights for our sample galaxies by averaging all the points within a radial bin of $0.1 \ r_{25}$ (red squares). The error bars on the squares represent 1$-\sigma$ scatter on the scale heights within a radial bin. We fit this mean normalized scale height with an exponential function of the form $h_{scl,n} = h_{0,n} \exp(R/l_{scl})$ and find an exponential scale length, $l_{scl} = (0.46 \pm 0.01) \ r_{25}$.

It is well known that the surface densities of the molecular gas in galaxies decline exponentially \citep{schruba11}. The scale length of this exponential decline (when normalised to 1 at 0.3 $r_{25}$) is found to be $l_{CO} = 0.20 \pm 0.01 \ r_{25}$ which is much less than the scale length of the molecular scale heights what we get here. The molecular surface density is expected to determine the molecular scale height directly in the absence of any external influence, such as the other baryonic discs or dark matter halo. In that case, the molecular scale height is expected to flare at the same rate as the molecular surface density declines. Subsequently, the scale length of the molecular surface density and the molecular scale height profile (HWHM) are expected to be the same. To test this, we modify Eq.~\ref{eq_hydro} to exclude the stellar disc, the atomic disc and the dark matter halo; and solve the resulting hydrostatic equation for a representative galaxy, NGC 3198. In Fig.~\ref{sclh_test} we plot the molecular surface density profile (solid red line) and the estimated molecular scale height (solid magenta line) as a function of radius. To estimate the scale lengths of these profiles, we fit both the curves with appropriate exponential functions (rising and falling). It should be emphasized here that the exponential fit to the molecular surface density profile is what we use as the input surface density to the hydrostatic equation. The molecular scale height profile is found to have a scale length of $\sim 4.06$ kpc which is almost the same as the scale length of the molecular disc which is $\sim 4.05$ kpc). However, for our sample galaxies, the scale length of the molecular scale height is found to be $\sim$ a factor of 2 greater than the scale length of the exponential surface density profile (see Fig.~\ref{sclh_norm}). This result strongly indicates that the coupling with the other disc components and with the dark matter halo significantly influence the molecular scale height.


Next, we use the solutions of the hydrostatic equilibrium equation and the observed rotation curves to build a dynamical model of the molecular discs for our sample galaxies. We then incline these dynamical models to the observed inclinations and project it to the sky-plane to produce a column density map. These maps then convolved with the telescope beam (13\arcsec $\times$ 13\arcsec) to produce simulated maps which are equivalent to the observed ones. In Fig.~\ref{ovrplt1} and~\ref{ovrplt2} we compare these simulated maps with the observed maps. In the left columns of the figures, we plot the modelled column density maps in contour on top of the observed maps in colour scale. As we do not solve the hydrostatic equation for the central regions of our sample galaxies, the central region should be excluded for any comparison. Further, the convolution with the telescope beam at the edge of the central region would corrupt an area equivalent to the beam size. We hence exclude a region which we do not solve from the centre plus the beam size. The vertical dashed lines in each panel of the left column represent this excluded region. As can be seen from the figure, the contours (modelled) match to the observed map reasonably well. However, for two galaxies, NGC 925 and NGC 2976 the model seems to deviate from the observation. This is due to azimuthal asymmetry in their column density distribution which is one of our implicit assumptions. We would like to emphasise here that, the molecular surface density profiles of our sample galaxies are calculated through spectral stacking method \citep{schruba11}. This produces surface density profiles which are much more sensitive than what will be observed in the moment zero maps. Because of this, the surface density profiles of our sample galaxies extend to a much larger radius as given in \citet{schruba11} than what can be found in \citet{leroy09a}. This also implies that our model molecular discs are expected to extend to a larger radius as opposed to what is apparent in the HERACLE survey maps. However, one advantage of our model molecular discs would then be, it can provide with a better estimation of the molecular discs for future sensitive observations.

As described in \S 1, the existence of thick molecular discs as observed through multiple observations are puzzling, and a lack of understanding persist regarding their origin and sustenance. As the molecular scale height is not a directly observable quantity in external galaxies, most of the strongest shreds of evidence come from indirect observation of the molecular gas velocity dispersion. To explore this further, we use our sample galaxies to examine how thick molecular discs they would produce while observed edge-on. This will also illuminate if a simple hydrostatic equilibrium condition can at all produce a thick molecular disc as observationally found. To do that, we use the dynamical models of the molecular discs of our sample galaxies and incline them to an inclination of 90$^o$ and generate model molecular edge-on column density maps. In the second columns of Fig.~\ref{ovrplt1} and~\ref{ovrplt2} we show the edge-on view of the molecular discs of our sample galaxies. To understand the edge-on thickness of our sample galaxies quantitatively, we extract the vertical column density profiles of the edge-on molecular discs and calculate the HWHM of it. This HWHM is an indicator of how thick a molecular disc is. In the right column of Fig.~\ref{ovrplt1} and~\ref{ovrplt2} we plot the edge-on HWHM profiles for our sample galaxies. 

As it can be seen from the figure, these HWHM profiles vary between $\sim 200-400$ pc at the central regions and can flare up to $\sim 400-500$ pc at the outer parts. This, in turn, indicates that at least our sample galaxies do produce a few kpc thick molecular disc when observed edge-on. \emph{Hence, it is very interesting that a thick molecular disc originates naturally under the assumption of a simple hydrostatic equilibrium.}


\section{Summary and future work}

In summary, we have selected a sample of eight galaxies from the HERACLE survey for which the surface density profiles of different baryonic discs, rotation curves and the mass models are available in the literature. Assuming a prevailing hydrostatic equilibrium in the baryonic discs under their mutual gravity and the external force field of the dark matter halo, we set up the joint Poisson's-Boltzman equation of hydrostatic equilibrium. Using an eight-order Runge-Kutta method, we solve the hydrostatic equation numerically for all our sample galaxies. 

The solutions of the hydrostatic equation are found to follow a $sech^2$ law in the absence of any coupling. However, it tends to behave more like a Gaussian than a $sech^2$ function in the presence of coupling. The solutions provide a three-dimensional distribution of the mass densities of different baryonic discs which we use to estimate the scale heights of the molecular and the atomic gas in our sample galaxies. We find that, in spite of having a wide range of surface density profiles, rotation curves and the dark matter halo structures, the molecular scale heights in our sample galaxies show very similar behaviour. It varies between $\sim 20-100$ pc at the central regions whereas it flares up to $\sim 150-200$ pc at the outer regions. The molecular scale heights found in our sample galaxies (typical spiral galaxies) are consistent with what is observed in the Milky Way. However, we also find that this molecular scale height considerably depends on the assumed $\sigma_{CO}$. A change of $\sim$ few \kms~in $\sigma_{CO}$ can lead to as much as $\sim 50\%$ change in the molecular scale height.  However, the $\sigma_{HI}$ found to have no meaningful effect on the molecular scale height.

We fit the molecular scale heights of our sample galaxies with an exponential profile and find that the molecular scale heights in our sample galaxies can be described very well by it. We further normalize the scale heights to investigate if the molecular scale heights follow a universal law or not. The normalized scale heights in our sample galaxies are found to follow a tight exponential law with very low scatter. The scale length of the exponential law is found to be $(0.46 \pm 0.01) \ r_{25}$. This scale length is found to be $\sim$ 2 times larger than the typical scale length of the molecular discs which indicates that, other disc environments play important role in deciding the molecular scale height in galaxies. 

We further use the density solutions and the rotation curves of the galaxies to build dynamical models of the molecular discs and consecutively produce model molecular column density maps. A comparison of these model maps with the real observations shows a reasonable agreement between the two. However, for two galaxies we find a considerable disagreement in the molecular column density maps because of their azimuthal asymmetry in the column density distribution and lack of sensitivity in the HERACLE data. 

We further use the dynamical models of the molecular discs and incline them to an inclination of 90$^o$ to examine the observed thickness of the molecular discs at an edge-on orientation. To quantitatively estimate the thickness of the molecular discs, we extract vertical molecular column density profiles and use them to estimate the HWHM profiles of the edge-on molecular discs. We find that for all our sample galaxies, the HWHM profiles vary between $\sim 300-500$ pc except for NGC 2976 for which we only solve out to a radius of $\sim$ 2 kpc and its HWHM reaches $\sim$ 200 pc which is roughly consistent with other galaxies at similar radii. This kind of HWHM values at edge-on orientation in our sample galaxies indicate that it is not difficult to produce a few kpc thick molecular disc in spiral galaxies just by assuming a prevailing hydrostatic equilibrium. 

However, as discussed earlier, in this study, we assumed our sample galaxies to have a single-component molecular disc with a single molecular velocity dispersion of 7 \kms. But as found in previous studies, there is a strong indication that the molecular discs in galaxies might have an extra low-density diffuse component which has a much higher scale height as reflected from their observed velocity dispersions. In these circumstances, one needs to add an extra thick molecular disc to the model of the galactic discs and subsequently modify the Poisson's-Boltzman equation. Not only that, as most of the previous studies concluded the existence of this diffuse molecular disc from spectral analysis, to understand it better from a theoretical perspective one needs to build spectral cubes and compare them to the observation which we plan to do next.

\bibliographystyle{mn2e}
\bibliography{bibliography}

\end{document}